\documentclass[10pt,aps,superscriptaddress,prc,eprint,nofootinbib,showkeys]{revtex4-2}
\pdfoutput=1
\usepackage{graphicx}
\usepackage{dcolumn}
\usepackage{bm}
\usepackage{amsmath,amsfonts,amsthm,amssymb}
\usepackage{graphics}
\usepackage{cancel}
\usepackage{multirow}
\usepackage{longtable}
\usepackage{color}
\usepackage[normalem]{ulem}
\usepackage{amssymb}
\usepackage{graphicx}
\usepackage{amsmath}
\usepackage{physics}
\usepackage{subfigure}
\usepackage{bbold}
\usepackage{yfonts}
\usepackage[colorlinks=true,linktocpage=true,linkcolor=blue,citecolor=blue]{hyperref}
\usepackage{placeins}
\usepackage{bm} 
\usepackage{nicefrac}
\usepackage{slashed}
\usepackage{marginnote}

\newcommand{\ab}[1]{\left\langle#1\right\rangle}

\begin{document}
	
\preprint{APS/123-QED}
	
\title{Relativistic spin hydrodynamics from novel relaxation time approximation}

\author{Samapan Bhadury}
\email{samapan.bhadury@uj.edu.pl}
\affiliation{Institute of Theoretical Physics, Jagiellonian University, ul. St. \L ojasiewicza 11, 30-348 Krakow, Poland}

\date{\today}

\begin{abstract}
    With the help of a semi-classical kinetic theory, a new collision kernel is proposed, which simultaneously conserves the energy-momentum tensor and the spin tensor of a relativistic fluid of spin-1/2 particles irrespective of the frame and matching conditions, even when relaxation time is momentum dependent. The relativistic Boltzmann's equation is solved using this new collision kernel to obtain the expressions of the transport coefficients with general definitions for the frame and matching conditions. The results indicate the expected existence of Barnett-like effect and the non-existence of Einstein--de-Haas-like effects. 
\end{abstract}

\maketitle

\section{Introduction}

During microscopic collisions, all fundamental interactions protect the conservation laws for the momentum four-vector ($p^\mu$) and total angular momentum tensor ($j^{\mu\nu} = \ell^{\mu\nu} + s^{\mu\nu}$), where, $\ell^{\mu\nu} = x^\mu p^\nu - x^\nu p^\mu$ is the orbital angular momentum with $x^\mu$ being the spacetime four-vector and $s^{\mu\nu}$ is the internal angular momentum. The latter was introduced by Mathisson \cite{Mathisson:1937zz} in 1937 and is related to spin four-vector ($s^{\mu}$) through the relation, $s^{\mu\nu} = \epsilon^{\mu\nu\alpha\beta} p_\alpha s_\beta/m$, where $m$ is the particle mass. Until recently, the study of relativistic kinetic theory focused on the microscopic conservation of the momentum four-vector, leading to the formulation of relativistic hydrodynamics while the angular momentum part was mostly ignored. With the discovery of spin polarization phenomena in heavy-ion collision experiments \cite{STAR:2017ckg, STAR:2018gyt, ALICE:2019aid, Liang:2004ph, Liang:2004xn, Becattini:2009wh, Becattini:2017gcx, STAR:2018pps, STAR:2018fqv, Mohanty:2021vbt, Palermo:2024tza}, the development of spin hydrodynamics from various approaches has gained significant interest \cite{Becattini:2013fla, Karabali:2014vla, Florkowski:2017ruc, Florkowski:2017dyn, Montenegro:2017lvf, Montenegro:2017rbu, Montenegro:2018bcf, Becattini:2018duy, Florkowski:2018fap, Florkowski:2019qdp, Montenegro:2020paq, Bhadury:2020puc, Becattini:2020sww, Shi:2020htn, Bhadury:2020cop, Fu:2020oxj, Speranza:2021bxf, Wang:2021ngp, Hongo:2021ona, Hu:2021pwh, Wang:2021wqq, Weickgenannt:2022zxs, Gallegos:2022jow, Sarwar:2022yzs, Biswas:2022bht, Biswas:2023qsw, Abboud:2023hos, Kiamari:2023fbe, Weickgenannt:2023nge, Ren:2024pur, Florkowski:2024bfw, Lin:2024cxo, Buzzegoli:2024mra, Fang:2024vds, Tiwari:2024trl, Fang:2024sym, Wagner:2024fry, Weickgenannt:2024ibf, Dey:2024cwo, She:2024rnx, Singh:2024cub, Huang:2024ffg, Daher:2025pfq}. Specifically, the non-trivial dynamics due to local \cite{anderson1974relativistic} and non local collisions \cite{Jaiswal:2012qm, Weickgenannt:2024ibf} causing exchange of spin in quantum kinetic theories are of particular interest \cite{Wagner:2023cct, Kumar:2023ojl}. In local collisions, the orbital and spin parts are conserved separately \cite{Florkowski:2018fap, Weickgenannt:2021cuo}. This simplification allows us to develop a theory of spin hydrodynamics for systems with conserved macroscopic spin tensor \cite{Bhadury:2020cop, Bhadury:2022ulr, Bhadury:2024ckc}. In general, such systems may be in an out-of-equilibrium state. The out-of-equilibrium dynamics is controlled by the collision kernel, whose complicated nature makes the Boltzmann equation an integro-differential equation. Consequently, some linearized approximations for collision kernel are used, which effectively preserve the properties of the original collision kernel. While the relaxation time approximation (RTA) used in Refs.~\cite{Bhadury:2020cop, Bhadury:2022ulr} fulfills the requirement of detailed balance of the collision kernel, it is not consistent with macroscopic conservation laws unless one imposes the specific Landau-Lifshitz (LL) frame and matching conditions \cite{landau1987fluid}. Recently it was shown in Ref.~\cite{Bhadury:2024ckc}, that the newly proposed collision kernel of extended relaxation time approximation (ERTA) \cite{Teaney:2013gca, Dash:2021ibx} can ensure all the conservation laws independent of the frame and matching conditions (even when the relaxation time is momentum and spin dependent) in an order-by-order manner in spacetime gradients. While the simplistic nature of ERTA is very useful \cite{Kamata:2022ola, Singh:2024leo}, it is desirable to construct a collision kernel, that ensures the conservation laws at all orders. This can be helpful for the development of a first-order causal and stable theory of spin hydrodynamics \cite{Bemfica:2017wps, Bemfica:2019cop, Kovtun:2019hdm, Bemfica:2019knx, Hoult:2020eho, Bemfica:2020xym, Bemfica:2020zjp, Hoult:2021gnb, Weickgenannt:2023btk} similar to the BDNK formalism of relativistic hydrodynamics for unpolarized systems.

In this work, we propose a new collision kernel following the development of novel relaxation time approximation (NRTA) \cite{Rocha:2021zcw, Rocha:2021lze} for a system of massive spin-half particles with zero chemical potential that are spin-polarizable. This collision kernel will be used to show that the macroscopic conservation laws of energy-momentum tensor ($T^{\mu\nu}$) and spin tensor ($S^{\lambda,\mu\nu}$) can be obtained at all orders as a direct consequence of the microscopic conservation laws, irrespective of any particular choice of frame and matching conditions. We will further solve the Boltzmann equation and obtain the transport coefficients associated with all the non-equilibrium processes. The results indicate the gradients of the fluid variables affect the evolution of spin transport, resembling the Barnett effect. However, the converse effect of Einstein--de-Haas is absent i.e. the evolution of fluid variables is independent of the gradients of macroscopic variables associated with spin.

The article is organized as follows: Section~\ref{sec:RSH} introduces the conserved currents of the system and their components. In Section~\ref{sec:RKT} the new collision kernel is proposed and the Boltzmann equation is solved. The transport coefficients are determined in Section~\ref{sec:TP} and we conclude the article with Section~\ref{sec:C&O}.

\textit{Notations and Conventions:} Throughout this article, natural units have been adopted i.e. $k_B = \hbar = c = 1$. For metric, the mostly negative convention has been chosen i.e., $g^{\mu\nu} = diag (1, -1, -1, -1)$ whereas for the Levi-Civita tensor the choice is, $\epsilon^{0123} = - \epsilon_{0123} = 1$. For both the ordinary phase-space distribution function, $f(x,p)$, and the extended phase-space distribution function, $f(x,p,s)$ the dependence on the microscopic variables are suppressed to denote them as, $f_{\textbf{p}}$ and $f_{s}$ respectively. Any macroscopic and microscopic variable at equilibrium will carry an additional subscript of `$0$' e.g., $f_{0\textbf{p}}$, whereas the non-equilibrium corrections will be denoted by adding a `$\delta$' in front of the variable e.g., $\delta f_{\textbf{p}}$. Bold letters will be used for three-vectors. The scalar product of four-vectors is denoted as, $A\cdot B \equiv A_\mu B^\mu$ and the scalar product of rank-2 tensors is denoted as, $A:B \equiv A_{\mu\nu} B^{\mu\nu}$.

\section{Relativistic Spin Hydrodynamics}
\label{sec:RSH}

For a rotating relativistic fluid of massive particles with vanishing chemical potential the relevant conserved currents are the energy-momentum tensor ($T^{\mu\nu}$), and the total angular momentum tensor ($J^{\lambda,\mu\nu}$), which is the sum of an orbital part ($L^{\lambda,\mu\nu} = x^\mu T^{\lambda\nu} - x^\nu T^{\lambda\mu}$) and a spin part ($S^{\lambda,\mu\nu}$) i.e. $J^{\lambda,\mu\nu} = L^{\lambda,\mu\nu} + S^{\lambda,\mu\nu}$. The non-uniqueness of the division of the orbital and spin parts of the angular momentum tensor gives rise to the problem of pseudogauge transformation. In the absence of non-local collisions, under the de-Groot--Leeuwen--Weert (GLW) pseudogauge, the energy-momentum tensor is symmetric \cite{DeGroot:1980dk} leading to a conserved spin tensor. The conserved currents for an out-of-equilibrium fluid can thus be expressed under GLW in terms of the single-particle distribution function in the extended phase-space as 
\begin{subequations}
    \begin{align}
        T^{\mu\nu} &= \varepsilon\, u^\mu u^\nu - P\, \Delta^{\mu\nu} + q^{\mu} u^{\nu} + q^{\nu} u^{\mu} + \pi^{\mu\nu} = \ab{p^\mu p^\nu}, \label{T^mn-decomp}\\
        S^{\lambda,\mu\nu} &= u^\lambda \mathcal{S}^{\mu\nu} + \left(u^{\mu} \Delta^{\nu\lambda} - u^{\nu} \Delta^{\mu\lambda}\right) \Sigma + u^\mu \Sigma^{\ab{\lambda\nu}}_{(s)} - u^\nu \Sigma^{\ab{\lambda\mu}}_{(s)} + u^\mu \Sigma_{(a)}^{\lambda\nu} - u^\nu \Sigma_{(a)}^{\lambda\mu} + \Sigma^{\lambda,\mu\nu} = \ab{p^\lambda s^{\mu\nu}}, \label{S^lmn-decomp}
    \end{align}
\end{subequations}
where, $\varepsilon$ and, $P$ are the full energy density and pressure respectively, which can be split into equilibrium and non-equilibrium parts as, $\varepsilon = \varepsilon_0 + \delta\varepsilon$ and, $P = P_0 + \delta P$ respectively, $u^\mu$ is the fluid four-velocity normalized to unity ($u_\mu u^\mu = 1$), the projection operator orthogonal to $u^\mu$ is defined as, $\Delta^{\mu\nu} = g^{\mu\nu} - u^{\mu} u^{\nu}$. Additionally, the currents, $q^\mu$ and $\pi^{\mu\nu}$ are the heat diffusion current and shear viscous pressure respectively. While the components of the spin tensors lack an unambiguous interpretation, in the present work, we will use the following terminologies, $\mathcal{S}^{\mu\nu}\to$ spin density, $\Sigma\to$ spin pressure, $\Sigma_{(s)}^{\ab{\mu\nu}}\to$ spin shear, $\Sigma_{(a)}^{\mu\nu}\to$ spin vorticity and, $\Sigma^{\lambda,\mu\nu}\to$ spin diffusion. Similar to $\varepsilon$, we can split these variables into equilibrium and non-equilibrium parts. Here the angular brackets enclosing the indices of the spin shear term are defined as, $A^{\langle\mu} B^{\nu\rangle} = \Delta^{\mu\nu}_{\alpha\beta} A^{\alpha} B^{\beta}$, where $\Delta^{\mu\nu}_{\alpha\beta} = \frac{1}{2} \left(\Delta^{\mu}_{\alpha} \Delta^{\nu}_{\beta} + \Delta^{\mu}_{\beta} \Delta^{\nu}_{\alpha}\right) - \frac{1}{3} \Delta^{\mu\nu} \Delta_{\alpha\beta}$ is the doubly symmetric traceless projection operator. The angular bracket appearing in Eqs.~\eqref{T^mn-decomp} and \eqref{S^lmn-decomp} is defined as,
\begin{align}
    \ab{\left(\cdots\right)} = \int_{\Gamma} \left(\cdots\right) f_{s}, \label{<>-def}
\end{align}
where $\int_{\Gamma} \equiv \int dP dS$ with $dP = \frac{g d^3\textbf{p}}{\left(2\pi\right)^3 E_{\textbf{p}}}$ being the momentum integral measure and, $dS = \frac{m}{\pi \mathfrak{s}} d^4 s \delta\left(s\cdot s + \mathfrak{s}^2\right) \delta \left(p\cdot s\right)$ being the spin integral measure. Here $g$ is the degeneracy factor (spin degeneracy not included), $E_{\textbf{p}} = \sqrt{\textbf{p}^2 + m^2}$ is the on-shell particle energy in the local rest frame and $\mathfrak{s}^2$ is the eigenvalue of the Casimir operator. Using the fact that the distribution function can be split into equilibrium and non-equilibrium parts as, $f_s = f_{0s} + \delta f_{s}$, one can also split the bracket in Eq.~\eqref{<>-def} as $\ab{\left(\cdots\right)} = \ab{\left(\cdots\right)}_0 + \ab{\left(\cdots\right)}_\delta$. One can intuitively understand the definition of these new brackets as, $\ab{\left(\cdots\right)}_{0} \equiv \int_\Gamma \left(\cdots\right) f_{0s}$ and, $\ab{\left(\cdots\right)}_{\delta} \equiv \int_\Gamma \left(\cdots\right) \delta f_{s}$. One more bracket will be used frequently in the article that is defined as, $\ab{\left(\cdots\right)}_{0\textbf{p}} \equiv \int dP \left(\cdots\right) f_{0\textbf{p}}$ The latter bracket will render the bracket $\ab{\left(\cdots\right)}_{\delta}$ redundant once $\delta f_s$ is defined in terms of $f_{0\textbf{p}}$.

The components of the conserved currents can now be expressed with the help of these brackets as,
\begin{subequations}
    \begin{align}
        \varepsilon_0 &= \ab{E_{\textbf{p}}^2}_{0},
        \quad\quad
        \delta \varepsilon = \ab{E_{\textbf{p}}^2}_{\delta},
        \quad\quad
        P_0 = - \ab{\left(1/3\right) \left(p\cdot\Delta\cdot p\right)}_{0},
        \quad\quad
        \delta P = - \ab{\left(1/3\right) \left(p\cdot\Delta\cdot p\right)}_{\delta}, \label{e,P-def}\\
        q^\mu &= \ab{E_{\textbf{p}} p^{\ab{\mu}}}_{\delta},
        \quad\quad
        \pi^{\mu\nu} = \ab{p^{\langle\mu} p^{\nu\rangle}}_{\delta}, \label{q^m,pi^mn-def}\\
        \mathcal{S}^{\mu\nu} &= u_\lambda S^{\lambda,\mu\nu} = \ab{E_{\textbf{p}} s^{\mu\nu}}
        = \ab{E_{\textbf{p}} s^{\mu\nu}}_{0} + \ab{E_{\textbf{p}} s^{\mu\nu}}_{\delta}, \label{spin-density}\\
        \Sigma &= \frac{1}{3} u_\mu \Delta_{\lambda\nu} S^{\lambda,\mu\nu} = \ab{\left(1/3\right) u_\mu p_{\ab{\nu}} s^{\mu\nu}} 
        %
        = 0, \label{spin-pressure}\\
        \Sigma_{(s)}^{\ab{\lambda\nu}} &= u_\mu \Delta^{\lambda\nu}_{\alpha\beta} S^{\alpha,\mu\beta} = \ab{u_\mu s^{\mu\langle\nu} p^{\lambda\rangle}}
        = \ab{u_\mu s^{\mu\langle\nu} p^{\lambda\rangle}}_{0} + \ab{u_\mu s^{\mu\langle\nu} p^{\lambda\rangle}}_{\delta}, \label{spin-shear}\\
        \Sigma_{(a)}^{\lambda\nu} &= u_\mu \Delta^{[\lambda}_\alpha \Delta^{\nu]}_\beta S^{\alpha,\mu\beta} = \ab{u_\mu s^{\mu[\nu} p^{\lambda]}}
        = \ab{u_\mu s^{\mu[\nu} p^{\lambda]}}_{0} + \ab{u_\mu s^{\mu[\nu} p^{\lambda]}}_{\delta}, \label{spin-vorticity}\\
        \Sigma^{\lambda,\mu\nu} &= \Delta^\lambda_\gamma \Delta^\mu_\alpha \Delta^\nu_\beta S^{\gamma,\alpha\beta} = \ab{p^{\ab{\lambda}} s^{\ab{\mu}\ab{\nu}}}
        = \ab{p^{\ab{\lambda}} s^{\ab{\mu}\ab{\nu}}}_{0} + \ab{p^{\ab{\lambda}} s^{\ab{\mu}\ab{\nu}}}_{\delta}, \label{spin-diffusion}
    \end{align}
\end{subequations}
where the notation $A^{\ab{\mu}} = \Delta^{\mu}_{\alpha} A^{\alpha}$ represents the part of the four-vector, $A^{\mu}$ that is orthogonal to $u_\mu$, $A^{[\mu} B^{\nu]} = \left(A^{\mu} B^{\nu} - A^{\nu} B^{\mu}\right)/2$ is the anti-symmetric combination. The spin integral measure enforces the Frenkel condition, $p_\mu s^{\mu\nu} = 0$, leading to a vanishing spin pressure. The terms with subscript `$0$' can be expressed in terms of $u^\mu$, temperature ($T$), and spin polarization tensor ($\omega^{\mu\nu}$), which is the Lagrange multiplier associated with the conservation of angular momentum, by noting that the equilibrium distribution can be defined,
\begin{align}
    f_{0s} = f_{0\textbf{p}} \exp\Big[\left(s:\omega\right)/2\Big], \label{f_0s-def}
\end{align}
where, $f_{0\textbf{p}} = e^{- \beta E_{\textbf{p}}}$. Here $\beta = 1/T$ is the inverse temperature. To determine the non-equilibrium quantities we need to know $\delta f_s \equiv \phi_s\, f_{0\textbf{p}}$. This can be obtained by solving the Boltzmann equation. In the next section, we will describe this process.

\section{Relativistic Kinetic Theory}
\label{sec:RKT}

\subsection{The Boltzmann equation and the collision kernel}
\label{ssec:Beq-CK}

The relativistic Boltzmann equation is given by,
\begin{align}
    p^\mu \partial_\mu f_{s} = C[f_s], \label{Beq1}
\end{align}
where $C[f_s]$ is the collision kernel. Thus, in order to solve the Boltzmann equation, we first need to specify the collision kernel. In case of $2\leftrightarrow2$ local collisions, considering only the cases where both momentum and spin are exchanged, with the help of the weak equivalence principle \cite{Weickgenannt:2020aaf}, the collision kernel can be shown to be given by \cite{Weickgenannt:2021cuo, Hu:2022lpi},
\begin{align}
    C[f_s] = \int_{\Gamma'} \int_{\Gamma_{1}} \int_{\Gamma'_{1}} W \left(f_{s_1} f_{s'_1} - f_{s} f_{s'}\right),\label{C[f]_22-def}
\end{align}
where $W$ is the transition matrix. Using the fact that we can write the distribution function as, $f_{s} = \left(1 + \phi_s\right) f_{0\textbf{p}}$ in Eq.~\eqref{C[f]_22-def} and keeping only the terms linear in $\phi_s$, we can re-write the Boltzmann equation with a linearized approximation of the collision kernel as,
\begin{align}
    p^\mu \partial_\mu f_s = \hat{L} \phi_s, \label{Beq2}
\end{align}
where the linearized collision operator, $\hat{L}$ can be interpreted as a linear operator in the Hilbert space, defined as,
\begin{align}
    \hat{L} \phi_s \equiv \int_{\Gamma'} \int_{\Gamma_{1}} \int_{\Gamma'_{1}} W f_{0 s} f_{0 s'} \left(\phi_{s_1} + \phi_{s'_1} - \phi_{s} - \phi_{s'}\right). \label{L-def}
\end{align}
Taking advantage of the self-adjointness property of such a linearized collision kernel and the fact that its eigenfunctions with zero eigenvalue correspond to the microscopically conserved degrees of freedom of the particle, the authors in Ref.~\cite{Rocha:2021lze} proposed a modified RTA such that one can obtain $\hat{L} 1 = 0$ and $\hat{L} k^\mu = 0$, which correspond to the microscopic conservation of particle number and linear momentum four-vector respectively. These properties of $\hat{L}$ allow one to build up a theory of relativistic hydrodynamics such that the number current and energy-momentum tensors are conserved by construction. Following this approach, in the present work, we propose a linearized collision kernel with the properties, $\hat{L} k^\mu = 0$ and, $\hat{L} s^{\mu\nu} = 0$, which will allow us to set up a theory of relativistic spin hydrodynamics\footnote{Note that, we do not demand $\hat{L} 1 = 0$. This is because such a constraint conserves the particle number separately and not the net particle number. In a relativistic system, where a particle and anti-particle pair can be produced, we can only demand the conservation of net particle number. However, this is a highly non-trivial task and is left for a future study. }. Consequently, in this case, the number of degenerate eigenfunctions ($\ket{\lambda_n}$) of $\hat{L}$ with zero eigenvalue is 10. Thus, the proposed abstract form of the linear collision operator in the Hilbert space is given by,
\begin{align}
    \hat{L} \sim - \mathbb{1} + \sum_{n=1}^{10} \ket{\lambda_n} \bra{\lambda_n}, \label{L_SNRTA-def}
\end{align}
where $\mathbb{1}$ corresponds to the Anderson-Witting RTA collision kernel. The new ten terms on the right-hand side correspond to counter terms that cancel out the homogeneous part of the solution of the Boltzmann equation. This cancellation allows one to build a theory of relativistic hydrodynamics independent of frame and matching conditions\footnote{In Refs.~\cite{Dash:2021ibx, Bhadury:2024ckc}, similar cancellations of the homogeneous part occur through the counter terms denoted by $\delta f^*$.}. The eigenfunctions, $\ket{\lambda_n}$ can be obtained explicitly with the help of Gram-Schmidt orthonormalization process. Following this process, the linear collision operator of Eq.~\eqref{L_SNRTA-def} can be expressed in terms of variables associated with spin hydrodynamics in the small polarization limit as,
\begin{align}
    &\hat{L} \phi_s = - \left(\frac{E_{\textbf{p}}}{\tau_{\rm R}}\right) f_{0s} \Bigg\{\phi_s - \frac{\ab{\left(E_{\textbf{p}}^2/\tau_{\rm R}\right) \phi_s}_0}{\ab{\left(E_{\textbf{p}}^3/\tau_{\rm R}\right)}_0} E_{\textbf{p}} - \frac{\ab{\left(E_{\textbf{p}}/\tau_{\rm R}\right) p^{\ab{\mu}} \phi_s}_0}{\ab{\left(1/3\right) \left(E_{\textbf{p}}/\tau_{\rm R}\right) p^{\ab{\alpha}} p_{\ab{\alpha}}}_0} p_{\ab{\mu}} \nonumber\\
    &- \bigg[\ab{\left(E_{\textbf{p}}/\tau_{\rm R}\right) \widetilde{s}_{\mu} \phi_s}_0 - \frac{\ab{\left(E_{\textbf{p}}^2/\tau_{\rm R}\right) \widetilde{s}_{\mu}}_0}{\ab{\left(E_{\textbf{p}}^3/\tau_{\rm R}\right)}_0} \ab{\left(E_{\textbf{p}}^2/\tau_{\rm R}\right) \phi_s}_0 - \frac{\ab{\left(E_{\textbf{p}}/\tau_{\rm R}\right) p^{\ab{\gamma}} \widetilde{s}_{\mu}}_0}{\ab{\left(1/3\right) \left(E_{\textbf{p}}/\tau_{\rm R}\right) p^{\ab{\alpha}} p_{\ab{\alpha}}}_0} \ab{\left(E_{\textbf{p}}/\tau_{\rm R}\right) p_{\ab{\gamma}} \phi_s}_0\bigg] \nonumber\\
    &\qquad\times \bigg[\widetilde{s}^{\,\mu} - \frac{\ab{\left(E_{\textbf{p}}^2/\tau_{\rm R}\right) \widetilde{s}^{\,\mu}}_0}{\ab{\left(E_{\textbf{p}}^3/\tau_{\rm R}\right)}_0} E_{\textbf{p}} - \frac{\ab{\left(E_{\textbf{p}}/\tau_{\rm R}\right) p^{\ab{\rho}} \widetilde{s}^{\,\mu}}_0}{\ab{\left(1/3\right) \left(E_{\textbf{p}}/\tau_{\rm R}\right) p^{\ab{\beta}} p_{\ab{\beta}}}_0} p_{\ab{\rho}}\bigg] \frac{1}{\ab{\left(1/3\right) \left(E_{\textbf{p}}/\tau_{\rm R}\right) \left(\widetilde{s} \cdot \widetilde{s}\,\right)}_0} \nonumber\\
    &- \bigg[\!\ab{\left(E_{\textbf{p}}/\tau_{\rm R}\right) \widetilde{s}_{\mu\nu} \phi_s}_0 \!-\! \frac{\ab{\left(E_{\textbf{p}}^2/\tau_{\rm R}\right) \widetilde{s}_{\mu\nu}}_0}{\ab{\left(E_{\textbf{p}}^3/\tau_{\rm R}\right)}_0} \ab{\left(E_{\textbf{p}}^2/\tau_{\rm R}\right) \phi_s}_0 \!-\! \frac{\ab{\left(E_{\textbf{p}}/\tau_{\rm R}\right) p^{\ab{\gamma}} \widetilde{s}_{\mu\nu}}_0}{\ab{\left(1/3\right) \left(E_{\textbf{p}}/\tau_{\rm R}\right) p^{\ab{\alpha}} p_{\ab{\alpha}}}_0} \ab{\left(E_{\textbf{p}}/\tau_{\rm R}\right) p_{\ab{\gamma}} \phi_s}_0 \!\bigg] \nonumber\\
    &\qquad\times \bigg[\widetilde{s}^{\,\mu\nu} - \frac{\ab{\left(E_{\textbf{p}}^2/\tau_{\rm R}\right) \widetilde{s}^{\,\mu\nu}}_0}{\ab{\left(E_{\textbf{p}}^3/\tau_{\rm R}\right)}_0} E_{\textbf{p}} - \frac{\ab{\left(E_{\textbf{p}}/\tau_{\rm R}\right) p^{\ab{\rho}} \widetilde{s}^{\,\mu\nu}}_0}{\ab{\left(1/3\right) \left(E_{\textbf{p}}/\tau_{\rm R}\right) p^{\ab{\beta}} p_{\ab{\beta}}}_0} p_{\ab{\rho}}\bigg] \frac{1}{\ab{\left(1/3\right) \left(E_{\textbf{p}}/\tau_{\rm R}\right) \left(\widetilde{s} : \widetilde{s}\,\right)}_0}\Bigg\}, \label{L_SNRTA-fin}
\end{align}
where $\widetilde{s}_\mu = u^{\beta} s_{\ab{\mu}\beta}$ and, $\widetilde{s}_{\mu\nu} = s_{\ab{\mu}\ab{\nu}}$. Here, the relaxation time could depend on particle momenta as well as spin. However, for the sake of simplicity, while determining the transport coefficients, the spin dependence on $\tau_{\rm R}$ has been ignored. Note that, all the terms appearing in Eq.~\eqref{L_SNRTA-fin} are essential to have $\hat{L} E_{\textbf{p}} = \hat{L} p^{\ab{\mu}} = \hat{L} s^{\mu\nu} = 0$ and hence the conservation laws i.e., $\partial_\mu T^{\mu\nu} = \partial_\mu \ab{p^{\mu} p^{\nu}}_0 = 0$, $\partial_\lambda s^{\lambda,\mu\nu} = \partial_\lambda \ab{p^{\lambda} S^{\mu\nu}}_0 = 0$ hold true for any order. Another crucial point is the distribution function appearing outside the curly braces in the first line of Eq.~\eqref{L_SNRTA-fin} is $f_{0s}$ and not $f_{0\textbf{p}}$. It ensures that the conservation laws are satisfied and can be understood by noting that the brackets, $\ab{\left(\cdots\right)}_{0}$ appearing in Eq.~\eqref{L_SNRTA-fin} are defined with the former. This complicates the solution process, but it is an unavoidable one.

\subsection{The out-of-equilibrium distribution function}
\label{ssec:OEDF}

Having introduced the desired collision kernel, the next step is to find the solution of $\phi_s$. The solution may be split into a spin-independent part and a spin-dependent part as \cite{Weickgenannt:2022zxs},
\begin{align}
    \phi_s = \phi_{\textbf{p}} + s:\psi_{\textbf{p}}, \label{phi_s-decomp}
\end{align}
Proceeding as in Refs.~\cite{Denicol:2012cn, Rocha:2021lze, Weickgenannt:2022zxs, Wagner:2024rbt} one may expand the solutions in terms of irreducible tensors as,
\begin{align}
    \phi_{\textbf{p}} = \sum_{n\in\mathbb{S}_0^{(\ell)}} \sum_{\ell=0}^\infty \Phi_n^{\ab{\mu_1\cdots\mu_\ell}} p_{\langle\mu_1} \cdots p_{\mu_\ell\rangle} P_n^{(0,\ell)},
    \quad\quad{\rm and,}\quad\quad
    \psi_{\textbf{p}}^{\mu\nu} = \sum_{n\in\mathbb{S}_1^{(\ell)}} \sum_{\ell=0}^\infty \Psi_n^{\mu\nu,\ab{\mu_1\cdots\mu_\ell}} p_{\langle\mu_1} \cdots p_{\mu_\ell\rangle} P_n^{(1,\ell)}, \label{phi_p,psi_p-expansion}
\end{align}
where, $p_{\langle\mu_1} \cdots p_{\mu_\ell\rangle} = \Delta^{\mu_1\cdots\mu_\ell}_{\nu_1\cdots\nu_\ell} p^{\langle\nu_1} \cdots p^{\nu_\ell\rangle}$ form the irreducible orthogonal momentum basis with the rank-$2\ell$ tensor $\Delta^{\mu_1\cdots\mu_\ell}_{\nu_1\cdots\nu_\ell}$ being defined in Refs,~\cite{DeGroot:1980dk, Denicol:2012cn}. Here, $\mathbb{S}_j^{(\ell)}$ represents the set of values that the index $n$ takes for a given value of $j$ and $\ell$. In principle, $\mathbb{S}_{j}^{(\ell)}$ should be the set, $\{0,1,2, \cdots \infty\}$. For analytical calculations, however, it often becomes necessary to truncate this set to finite values\footnote{In the present work, it is still possible to solve for $\phi_{\textbf{p}}$ without truncating the set $\mathbb{S}_0^{(\ell)}$ following the Appendix A of Ref.~\cite{Rocha:2022fqz}. However, this will make it impossible to obtain an analytical solution for $\psi_{\textbf{p}}^{\mu\nu}$.}. In this work, we have chosen the minimal set necessary to construct the theory of spin hydrodynamics with general frame and matching conditions. These are, $\mathbb{S}_0^{(0)} = \{0,1,2\}$, $\mathbb{S}_0^{(1)} = \{0,1\}$, $\mathbb{S}_0^{(2)} = \{0\}$ and, $\mathbb{S}_1^{(0)} = \{0,1\}$, $\mathbb{S}_1^{(1)} = \{0\}$, while all other $\mathbb{S}_j^{(\ell)}$ are null sets. This implies, there are $14+30=44$ degrees of freedom. However, by imposing the frame and matching conditions, the degrees of freedom are further reduced by $4+6=10$ leaving us with $10+24=34$ degrees of freedom in total. However, of these, $4$ spin degrees of freedom are redundant due to $p_\mu s^{\mu\nu} = 0$. The expansions in Eq.~\eqref{phi_p,psi_p-expansion}, require one to use the set of orthogonal polynomials of energy, $P_n^{(j,\ell)}$ satisfying the orthogonality condition,
\begin{align}
    \frac{\ell!}{\left(2\ell+1\right)!!} \int dP \left(E_{\textbf{p}}/\tau_{\rm R}\right) \left(p\cdot\Delta\cdot p\right)^\ell P_m^{(j,\ell)} P_n^{(j,\ell)} f_{0\textbf{p}} = A_n^{(j,\ell)} \delta_{mn}. \label{<PP>-ortho-n}
\end{align}
where,
\begin{align}
    A_{n}^{(j,\ell)} = \frac{\ell!}{\left(2\ell+1\right)!!} \ab{\left(E_{\textbf{p}}/\tau_{\rm R}\right) \left(p\cdot\Delta\cdot p\right)^\ell P_{n}^{(j,\ell)} P_{n}^{(j,\ell)}}_{0\textbf{p}}. \label{A_n^(j,l)}
\end{align}
The polynomials $P_n^{(j,\ell)}$ can be determined through the Gram-Schmidt process. However, one has to still define one element of the set to initiate the process. In the present work, we impose, $P_0^{(0,0)} = \beta E_{\textbf{p}}$ and, $P_0^{(0,\ell\geq1)} = P_0^{(1,\ell\geq0)} = 1$. The expansion coefficients, $\Phi_n^{\ab{\mu_1\cdots\mu_\ell}}$ and $\Psi_n^{\mu\nu,\ab{\mu_1\cdots\mu_\ell}}$ are functions of macroscopic variables only. Furthermore, $\Psi_n^{\mu\nu,\ab{\mu_1\cdots\mu_\ell}} \sim \mathcal{O} \left(\omega\right)$ and hence, we will ignore terms that go like $\psi_{\textbf{p}} \omega^{n}$ for $n\geq1$. Note that, here the index $\ell$ corresponds to the tensorial rank of the coefficient and $n$ represents the $n$-th coefficient of the $\ell$-th rank coefficient. In the present work, $j=0$ and $j=1$ correspond to the spin-independent and spin-dependent parts respectively.

Therefore, all that remains is to determine coefficients, $\Phi_n^{\ab{\mu_1\cdots\mu_\ell}}$ and, $\Psi_n^{\mu\nu,\ab{\mu_1\cdots\mu_\ell}}$. To obtain these coefficients up to first order in spacetime gradients, we first consider the left-hand side of Eq.~\eqref{Beq2}. Following the Chapman-Enskog-like expansion, only the first-order gradient terms are kept so that the linearized Boltzmann equation is given by,
\begin{align}
    \left(A_{\textbf{p}} + B_{\textbf{p}}^{\ab{\mu}}\, p_\mu + C_{\textbf{p}}^{\ab{\mu\nu}} p_\mu p_\nu\right) f_{0\textbf{p}} \left[1 + \frac{1}{2} \left(s:\omega\right)\right] + D_{\textbf{p}}^{\mu,\alpha\beta} p_\mu s_{\alpha\beta} f_{0\textbf{p}} = \hat{L} \phi_s, \label{Beq3}
\end{align}
where,
\begin{subequations}
    \begin{align}
        A_{\textbf{p}} &= - E_{\textbf{p}}^2\, \Dot{\beta} - \frac{\beta}{3} \left(p\cdot\Delta\cdot p\right) \theta,
        \qquad\qquad
        B_{\textbf{p}}^{\ab{\mu}} = - E_{\textbf{p}} \left(\nabla^\mu \beta\right) - \beta E_{\textbf{p}} \Dot{u}^\mu, \label{A_p;B_p^m-def}\\
        C_{\textbf{p}}^{\ab{\mu\nu}} &= - \beta \sigma^{\mu\nu},
        \hspace{3.55cm}
        D_{\textbf{p}}^{\mu,\alpha\beta} = \frac{1}{2} \left(u^\mu \Dot{\omega}^{\alpha\beta} + \nabla^\mu \omega^{\alpha\beta}\right). \label{C_p^mn;D_s^m,ab-def}
    \end{align}
\end{subequations}
where, $\theta \equiv \partial\cdot u$ is the expansion scalar and, $\sigma^{\mu\nu} = \partial^{\langle\mu} u^{\nu\rangle}$ is the velocity stress tensor, $\Dot{A} \equiv \left(u\cdot \partial\right) A$ denotes the co-moving derivative, $\nabla^\mu \equiv \Delta^{\mu}_{\alpha} \partial^{\alpha}$ denotes the spacelike derivative. On the right-hand side of Eq.~\eqref{Beq2}, one may use Eq.~\eqref{L_SNRTA-fin} to substitute the linear collision kernel and $\phi_s$ is substituted from Eqs.~\eqref{phi_s-decomp} and \eqref{phi_p,psi_p-expansion}. However, not all of the coefficients can be determined from the Boltzmann equation. The ones corresponding to homogeneous solution i.e. $\Phi_0$, $\Phi_{1}^{\ab{\mu}}$, and $\Psi_{0}^{\mu\nu}$ are obtained through the frame and matching conditions which in the present case are taken to be,
\begin{align}
    &\int_\Gamma q_1\, \phi_s\, f_{0\textbf{p}} = 0,
    \qquad\qquad
    \int_\Gamma q_2\, p^{\ab{\mu}} \, \phi_s\, f_{0\textbf{p}}= 0,
    \qquad\qquad
    \int_\Gamma q_3\, s^{\mu\nu} \phi_s\, f_{0\textbf{p}} = 0, \label{F&M-conds}
\end{align}
where, $q_{1/2/3}$ are not specified at the moment. The choice, $q_1 = E_{\textbf{p}}^2$, and $q_2 = q_3 = E_{\textbf{p}}$ lead to the ones used in Refs.~\cite{Bhadury:2020cop, Bhadury:2022ulr}. The solutions can be now expressed as,
\begin{subequations}
    \begin{align}
        \phi_{\textbf{p}} = \sum_{n\in\mathbb{S}_0^{(0)}} \Phi_n P_n^{(0,0)} + \sum_{n\in\mathbb{S}_0^{(1)}} \Phi_n^{\ab{\mu_1}} P_n^{(0,1)} p_{\ab{\mu_1}} + \sum_{n\in\mathbb{S}_0^{(2)}} \Phi_n^{\ab{\mu_1\mu_2}} P_n^{(0,2)} p_{\langle\mu_1} p_{\mu_2\rangle}, \label{phi_p-expr}
    \end{align}
    \begin{align}
        \psi_{\textbf{p}}^{\mu\nu} = \sum_{n\in\mathbb{S}_1^{(0)}} \Psi_n^{\mu\nu} P_n^{(1,0)} + \sum_{n\in\mathbb{S}_1^{(1)}} \Phi_n^{\mu\nu,\ab{\mu_1}} p_{\ab{\mu_1}} P_n^{(1,1)}. \label{psi_p-expr}
    \end{align}
\end{subequations}
where,
\begin{subequations}
    \begin{align}
        \Phi_0 &= - \sum_{n\in\mathbb{S}_0^{(0)} -\, \{0\}} \frac{\ab{q_1\, P_n^{(0,0)}}_{0\textbf{p}}}{\ab{q_1\, P_0^{(0,0)}}_{0\textbf{p}}} \Phi_n,
        \qquad\qquad
        \Phi_n = - \frac{\ab{A_{\textbf{p}} P_n^{(0,0)}}_{0\textbf{p}}}{A_n^{(0,0)}}, \qquad\qquad{\rm for~}n=1,2 \label{Phi_n-expr}\\
        \Phi_0^{\ab{\mu}} &= - \frac{\ab{q_2 \left(p \cdot \Delta \cdot p\right) P_1^{(0,1)}}_{0\textbf{p}}}{\ab{q_2 \left(p\cdot\Delta\cdot p\right) P_0^{(0,1)}}_{0\textbf{p}}} \Phi_1^{\ab{\mu}}
        \qquad\qquad
        \Phi_1^{\ab{\mu}} = - \frac{1}{A_1^{(0,1))}} \ab{B_{\textbf{p}}^{\ab{\mu}} \left(p\cdot\Delta\cdot p\right) P_1^{(0,1)}}_{0\textbf{p}}, \label{Phi_n^m-expr} \\
        \Phi_0^{\langle\mu\nu\rangle} &= - \frac{1}{A_{0}^{(0,2)}} \ab{C_{\textbf{p}}^{\ab{\mu\nu}} \left(p\cdot\Delta\cdot p\right)^2 P_0^{(0,2)}}_{0\textbf{p}}, \label{Phi_0^mn-expr}
    \end{align}
    \begin{align}
        \Psi_{0}^{\mu\nu} &= - \left[\frac{\ab{q_3 P_{1}^{(1,0)}}_{0\textbf{p}}}{\ab{q_3 P_{0}^{(1,0)}}_{0\textbf{p}}}\right] \Psi_{1}^{\mu\nu},
        \qquad\qquad
        \Psi_{1}^{\mu\nu} = C_1^{\mu\nu},
        \qquad\qquad
        \Psi_{0}^{\mu\nu,\ab{\mu_1}} = C_{0}^{\mu\nu, \ab{\mu_1}}, \label{Psi_{0,1}^mn,<>-expr}
    \end{align}
\end{subequations}
with $C_1^{\mu\nu}$ being the solution of the equation,
\begin{align}
    C_{1}^{\mu\nu} + A_1^{\mu\nu\alpha\beta} C_{1,\alpha\beta} = B_1^{\mu\nu}. \label{C_1^mn-eq}
\end{align}
The expressions of $A_1^{\mu\nu\alpha\beta}$, $B_1^{\mu\nu}$ and, $C_{0}^{\mu\nu, \ab{\mu_1}}$ are given in Appendix~\ref{app:A}. It is possible to solve the equation of the form Eq.~\eqref{C_1^mn-eq} by constructing projection operators as discussed in Ref.~\cite{Hess:2015szz} (see Ref.~\cite{Panda:2020zhr} for an example of the solution from magnetohydrodynamics). It may be noted that, in Eq.~\eqref{L_SNRTA-fin} while all of the 21 terms are necessary for conservation laws, only 5 of them contribute in the determination of the coefficients $\Phi_n^{\ab{\mu_1\cdots\mu_\ell}}$ and $\Psi_n^{\mu\nu,\ab{\mu_1\cdots\mu_\ell}}$. Another important point is, the coefficients of $\psi_{\textbf{p}}^{\mu\nu}$ were obtained using the fact that due to the presence of the term, $s:\psi_{\textbf{p}}$ in Eq.~\eqref{phi_s-decomp}, we can ignore any part of $\psi_{\textbf{p}}^{\mu\nu}$ that is parallel to momentum four-vector i.e. we have, $p_\mu \psi_{\textbf{p}}^{\mu\nu} = 0 = p_\mu \psi_{\textbf{p}}^{\nu\mu}$.

\section{Transport Properties}
\label{sec:TP}

In the previous section, the non-equilibrium corrections to the distribution function have been determined. Using this information, the transport properties can now be evaluated. Substituting Eq.~\eqref{phi_s-decomp} with the help of Eqs.~\eqref{phi_p-expr}, \eqref{psi_p-expr} and, \eqref{Phi_n-expr}-\eqref{Psi_{0,1}^mn,<>-expr} into Eqs.~\eqref{e,P-def}-\eqref{spin-diffusion} one finds,
\begin{subequations}
    \begin{align}
        \delta \varepsilon &= \sum_{n\in\mathbb{S}_0^{(0)}} \Phi_{n} \ab{E_{\textbf{p}}^2 P_n^{(0,0)}}_{0\textbf{p}},
        \hspace{4cm}
        \delta P = - \sum_{n\in\mathbb{S}_0^{(0)}} \Phi_{n} \ab{\left(1/3\right) \left(p\cdot\Delta\cdot p\right) P_n^{(0,0)}}_{0\textbf{p}}, \label{TC:de,dP}\\
        q^\mu &= \sum_{n\in\mathbb{S}_0^{(1)}} \Phi_{n}^{\ab{\mu_1}} \ab{\left(1/3\right) E_{\textbf{p}} \left(p\cdot\Delta\cdot p\right) P_n^{(0,1)}}_{0\textbf{p}},
        \hspace{1.1cm}
        \pi^{\mu\nu} = \Phi_{0}^{\ab{\mu\nu}} \ab{\left(2/15\right) \left(p\cdot\Delta\cdot p\right)^2}_{0\textbf{p}}, \label{TC:q^m,pi^mn}\\
        \delta \mathcal{S}^{\mu\nu} &= \sum_{n\in\mathbb{S}_1^{(0)}} \Psi_{n}^{\mu\nu} \ab{E_{\textbf{p}} P_n^{(1,0)}}_{0\textbf{p}},
        \hspace{3.25cm}
        \delta \Sigma_{(s)}^{\ab{\mu\nu}} = u_\alpha \Delta^{\mu\nu}_{\rho\gamma} \Psi_{0}^{\alpha\rho,\ab{\gamma}} \ab{\left(1/3\right) \left(p\cdot\Delta\cdot p\right)}_{0\textbf{p}}, \label{TC:dS^mn,SS}\\
        \delta \Sigma_{(a)}^{\mu\nu} &= \sum_{n\in\mathbb{S}_1^{(0)}}\!\! u_\alpha \Psi_{n}^{\alpha[\mu} u^{\nu]} \ab{E_{\textbf{p}} P_n^{(1,0)}}_{\!0\textbf{p}} \!\!+\! u_\alpha \Psi_{0}^{\alpha[\mu,\ab{\nu}]} \ab{\left(1/3\right) \left(p\cdot\Delta\cdot p\right)}_{0\textbf{p}},
        \hspace{0.5cm}
        \delta \Sigma^{\lambda,\mu\nu} = \Psi_{0}^{\ab{\mu}\ab{\nu},\ab{\lambda}} \ab{\left(p\cdot\Delta\cdot p\right)}_{\!0\textbf{p}}, \label{TC:SD}
    \end{align}
\end{subequations}
If one assumes the relaxation time to be independent of particle momenta and uses $q_1 = E_{\textbf{p}}^2$, $q_2 = q_3 = E_{\textbf{p}}$ in Eq.~\eqref{F&M-conds}, then the collision kernel reduces back to the standard RTA \cite{anderson1974relativistic} and the transport coefficients coincide with those obtained in Ref.~\cite{Bhadury:2020cop}.
The expressions of Eqs.~\eqref{TC:de,dP}-\eqref{TC:q^m,pi^mn} show that the hydrodynamic variables remain impervious to the spin dynamics, whereas the spin transport properties are influenced by the evolution of hydrodynamic variables. This is the expected behavior of relativistic fluids at small polarization in the absence of any external field's influence, as observed in Refs.~\cite{Bhadury:2020cop, Bhadury:2022ulr, Bhadury:2024ckc}. Such phenomenon is analogous to Barnett effect \cite{Barnett:1935wyv} where a rotating magnetizable fluid can become magnetized under the influence of the rotation.

\section{Conclusions and outlook}
\label{sec:C&O}

In this work, we proposed a new collision kernel that consistently guarantees macroscopic conservation laws of energy-momentum tensor and the spin tensor of a relativistic fluid whose evolution is dictated by local collisions. We solved the Boltzmann equation with this collision kernel in the limit of small polarization such that the relaxation time can be a function of particle four-momenta. The transport coefficients were determined under general frame and matching conditions. Following the BDNK formulation of first order relativistic hydrodynamics, the approach described here does not replace the co-moving derivatives. One may therefore expect this theory of first-order spin hydrodynamics to be causal. Hence, in future, we would like to perform the linear mode analysis of this theory.

We would like to extend this formulation to include the effect of the background magnetic field and obtain the transport coefficients for spin-magnetohydrodynamics \cite{Bhadury:2022ulr}. However, this can be achieved only after deriving a collision kernel that can describe particle and anti-particle pair production and annihilation. In the present work, the part of the collision kernel that exchanges only spin has been ignored for the sake of simplicity. It will be interesting to explore the implications of such a term in future studies.

\section*{Acknowledgment}

I am highly grateful to Gabriel S. Rocha and David Wagner for the useful correspondence and discussions. I would like to thank Prof. Amaresh Jaiswal for the valuable feedback on the manuscript. I acknowledge the support of the Faculty of Physics, Astronomy and Applied Computer Science, Jagiellonian University Grant No. LM/36/BS.

\appendix

\section{List of Coefficients}
\label{app:A}

Here we provide the details of the coefficients, $A_{1}^{\mu\nu\alpha\beta}, B_{1}^{\mu\nu}$ and $C_{0}^{\mu\nu,\ab{\mu_1}}$ used in Section~\ref{sec:RKT}.
\begin{align}
    A_{1}^{\mu\nu\alpha\beta} &= \left\{\frac{\ab{\left(2/3\right) \left(E_{\textbf{p}}/\tau_{\rm R}\right) \left(p\cdot\Delta\cdot p\right) P_1^{(1,0)}}_{0\textbf{p}}}{\ab{\left(1/3\right) \left(E_{\textbf{p}}/\tau_{\rm R}\right) \left(p\cdot\Delta\cdot p\right)}_{0\textbf{p}}} \Delta_{\alpha[\mu} u_{\nu]} u_{\beta}\right. \nonumber\\
    &\qquad\left.+ \frac{\left[\ab{\left(E_{\textbf{p}}^3/\tau_{\rm R}\right) P_1^{(1,0)}}_{0\textbf{p}} + \ab{\left(1/3\right) \left(E_{\textbf{p}}/\tau_{\rm R}\right) \left(p\cdot\Delta\cdot p\right) P_1^{(1,0)}}_{0\textbf{p}}\right]}{\left[\ab{\left(E_{\textbf{p}}^3/\tau_{\rm R}\right)}_{0\textbf{p}} + \ab{\left(1/3\right) \left(E_{\textbf{p}}/\tau_{\rm R}\right) \left(p\cdot\Delta\cdot p\right)}_{0\textbf{p}}\right]} \Delta_{\mu\alpha} \Delta_{\nu\beta}\right\} \frac{\ab{q_3 P_{1}^{(1,0)}}_{0\textbf{p}}}{\ab{q_3 P_{0}^{(1,0)}}_{0\textbf{p}}} \frac{A_{0}^{(1,0)}}{A_{1}^{(1,0)}}, \label{A_1^mnab}\\
    B_{1}^{\mu\nu} &= - \frac{1}{A_{1}^{(1,0)}} \ab{P_{1}^{(1,0)} s^{\mu\nu} \left(p\cdot\partial\right)}_{0} - \frac{\ab{\left(E_{\textbf{p}}/\tau_{\rm R}\right) \phi_\textbf{p} P_1^{(1,0)} s^{\mu\nu}}_{0}}{A_{1}^{(1,0)}} + \frac{\ab{\left(E_{\textbf{p}}/\tau_{\rm R}\right) E_{\textbf{p}} P_1^{(1,0)} s^{\mu\nu}}_{0}}{A_{1}^{(1,0)}} \sum_{n\in\mathbb{S}_0^{(0)}} \Phi_n \frac{\ab{\left(E_{\textbf{p}}^2/\tau_{\rm R}\right) P_n^{(0,0)}}_{0\textbf{p}}}{\ab{\!\left(E_{\textbf{p}}^3/\tau_{\rm R}\right)}_{0\textbf{p}}} \nonumber\\
    &\qquad+ \frac{\ab{\left(E_{\textbf{p}}/\tau_{\rm R}\right) p_{\ab{\mu}} P_1^{(1,0)} s^{\mu\nu}}_{0}}{A_{1}^{(1,0)}} \sum_{n\in\mathbb{S}_0^{(1)}} \Phi_n^{\ab{\mu}} \frac{\ab{\left(1/3\right) \left(E_{\textbf{p}}/\tau_{\rm R}\right) \left(p\cdot\Delta\cdot p\right) P_n^{(0,1)}}_0}{\ab{\left(1/3\right) \left(E_{\textbf{p}}/\tau_{\rm R}\right) \left(p\cdot\Delta\cdot p\right)}_0} \nonumber\\
    &\qquad+ 2 \Delta^{\mu}_{[\gamma} u_{\lambda]} \frac{\ab{\left(1/3\right) \left(E_{\textbf{p}}/\tau_{\rm R}\right) \left(p\cdot\Delta\cdot p\right) P_1^{(1,0)}}_{0\textbf{p}}}{\ab{\left(1/3\right) \left(E_{\textbf{p}}/\tau_{\rm R}\right) \left(p\cdot\Delta\cdot p\right)}_{0\textbf{p}}} \frac{\ab{\left(E_{\textbf{p}}/\tau_{\rm R}\right) \widetilde{s}_{\mu} \phi_{\textbf{p}}}_{0}}{A_{1}^{(1,0)}} \nonumber\\
    &\qquad+ \frac{\left[\ab{\left(E_{\textbf{p}}^3/\tau_{\rm R}\right) P_1^{(1,0)}}_{0\textbf{p}} + \ab{\left(1/3\right) \left(E_{\textbf{p}}/\tau_{\rm R}\right) \left(p\cdot\Delta\cdot p\right) P_1^{(1,0)}}_{0\textbf{p}}\right]}{\left[\ab{\left(E_{\textbf{p}}^3/\tau_{\rm R}\right)}_{0\textbf{p}} + \ab{\left(1/3\right) \left(E_{\textbf{p}}/\tau_{\rm R}\right) \left(p\cdot\Delta\cdot p\right)}_{0\textbf{p}}\right]} \frac{\ab{\left(E_{\textbf{p}}/\tau_{\rm R}\right) \widetilde{s}^{\mu\nu} \phi_{\textbf{p}}}_0}{A_{1}^{(1,0)}}, \label{B_1^mn}\\
    C_0^{\mu\nu,\ab{\mu_1}} &=\! \frac{1}{A_0^{(1,1)}} \Bigg\{\!- \ab{P_{0}^{(1,1)} p^{\ab{\mu_1}} s^{\mu\nu} \left(p\cdot\partial\right)}_{0} \!-\! \ab{\!\left(E_{\textbf{p}}/\tau_{\rm R}\right) \phi_\textbf{p} P_{0}^{(1,1)} s^{\mu\nu}}_{0} \!+ \ab{\!\left(E_{\textbf{p}}^2/\tau_{\rm R}\right) p^{\ab{\mu_1}} P_{0}^{(1,1)} s^{\mu\nu}}_{0} \!\!\!\sum_{n\in\mathbb{S}_0^{(0)}}\!\! \Phi_n \frac{\ab{\!\left(E_{\textbf{p}}^2/\tau_{\rm R}\right) P_n^{(0,0)}}_{0}}{\ab{\left(E_{\textbf{p}}^3/\tau_{\rm R}\right)}_{0}} \nonumber\\
    + &\ab{\!\left(E_{\textbf{p}}^2/\tau_{\rm R}\right) p_{\ab{\gamma}} p^{\ab{\mu_1}} P_{0}^{(1,1)} s^{\mu\nu}}_{0} \!\!\!\sum_{n\in\mathbb{S}_0^{(1)}} \!\!\!\Phi_n^{\ab{\gamma}} \frac{\ab{\left(1/3\right) \left(E_{\textbf{p}}/\tau_{\rm R}\right) \left(p\cdot\Delta\cdot p\right) P_n^{(0,1)}}_{0}}{\ab{\left(1/3\right) \left(E_{\textbf{p}}/\tau_{\rm R}\right) \left(p\cdot\Delta\cdot p\right)}_{0}} \nonumber\\
    + &\Delta^{\gamma[\mu} \Delta^{\nu]\mu_1} \ab{\left(1/3\right) \left(E_{\textbf{p}}^2/\tau_{\rm R}\right) \left(p\cdot\Delta\cdot p\right) P_{0}^{(1,1)}}_{0\textbf{p}} \left[\frac{\ab{\left(E_{\textbf{p}}/\tau_{\rm R}\right) \widetilde{s}_{\gamma} \phi_{\textbf{p}}}_{0}}{\ab{\left(1/3\right) \left(E_{\textbf{p}}/\tau_{\rm R}\right) \left(p\cdot\Delta\cdot p\right)}_{0\textbf{p}}} \!+\! \sum_{n\in\mathbb{S}_1^{(0)}} \widetilde{\Psi}_{n,\gamma} \frac{\ab{\left(E_{\textbf{p}}/\tau_{\rm R}\right) P_n^{(1,0)}}_{0\textbf{p}}}{\ab{\left(1/3\right) \left(E_{\textbf{p}}/\tau_{\rm R}\right) \left(p\cdot\Delta\cdot p\right)}_{0\textbf{p}}}\right] \nonumber\\
    + &\frac{2 \ab{\left(1/3\right) \left(E_{\textbf{p}}^2/\tau_{\rm R}\right) P_{0}^{(1,1)} \left(p\cdot\Delta\cdot p\right)}_{0\textbf{p}}}{\left[\ab{\left(E_{\textbf{p}}^3/\tau_{\rm R}\right)}_{0\textbf{p}} + \ab{\left(1/3\right) \left(E_{\textbf{p}}/\tau_{\rm R}\right) \left(p\cdot\Delta\cdot p\right)}_{0\textbf{p}}\right]} u^{[\mu} \Delta^{\nu][\gamma} \Delta^{\lambda]\mu_1} \bigg[\ab{\left(E_{\textbf{p}}/\tau_{\rm R}\right) \widetilde{s}_{\gamma\lambda} \phi_{\textbf{p}}}_0 + \sum_{n\in\mathbb{S}_1^{(0)}} \widetilde{\Psi}_{n,\gamma\lambda} \ab{\left(E_{\textbf{p}}/\tau_{\rm R}\right) P_n^{(1,0)}}_{0\textbf{p}} \bigg]\Bigg\}. \label{C_0^mn<m1>}
\end{align}
Note, for some function, $g(x,\!p,\!s)$, we defined the quantity, $\ab{g(x,p,s) \left(p\cdot \partial\right)}_{0} \!=\! \int_\Gamma g(x,p,s) \left(p\cdot \partial\right)\! f_{0s}$ in Eqs.~\eqref{A_1^mnab}-\eqref{C_0^mn<m1>}.

\bibliography{ref}

\begin{thebibliography}{87}%
\makeatletter
\providecommand \@ifxundefined [1]{%
 \@ifx{#1\undefined}
}%
\providecommand \@ifnum [1]{%
 \ifnum #1\expandafter \@firstoftwo
 \else \expandafter \@secondoftwo
 \fi
}%
\providecommand \@ifx [1]{%
 \ifx #1\expandafter \@firstoftwo
 \else \expandafter \@secondoftwo
 \fi
}%
\providecommand \natexlab [1]{#1}%
\providecommand \enquote  [1]{``#1''}%
\providecommand \bibnamefont  [1]{#1}%
\providecommand \bibfnamefont [1]{#1}%
\providecommand \citenamefont [1]{#1}%
\providecommand \href@noop [0]{\@secondoftwo}%
\providecommand \href [0]{\begingroup \@sanitize@url \@href}%
\providecommand \@href[1]{\@@startlink{#1}\@@href}%
\providecommand \@@href[1]{\endgroup#1\@@endlink}%
\providecommand \@sanitize@url [0]{\catcode `\\12\catcode `\$12\catcode `\&12\catcode `\#12\catcode `\^12\catcode `\_12\catcode `\%12\relax}%
\providecommand \@@startlink[1]{}%
\providecommand \@@endlink[0]{}%
\providecommand \url  [0]{\begingroup\@sanitize@url \@url }%
\providecommand \@url [1]{\endgroup\@href {#1}{\urlprefix }}%
\providecommand \urlprefix  [0]{URL }%
\providecommand \Eprint [0]{\href }%
\providecommand \doibase [0]{http://dx.doi.org/}%
\providecommand \selectlanguage [0]{\@gobble}%
\providecommand \bibinfo  [0]{\@secondoftwo}%
\providecommand \bibfield  [0]{\@secondoftwo}%
\providecommand \translation [1]{[#1]}%
\providecommand \BibitemOpen [0]{}%
\providecommand \bibitemStop [0]{}%
\providecommand \bibitemNoStop [0]{.\EOS\space}%
\providecommand \EOS [0]{\spacefactor3000\relax}%
\providecommand \BibitemShut  [1]{\csname bibitem#1\endcsname}%
\let\auto@bib@innerbib\@empty
\bibitem [{\citenamefont {Mathisson}(1937)}]{Mathisson:1937zz}%
  \BibitemOpen
  \bibfield  {author} {\bibinfo {author} {\bibfnamefont {M.}~\bibnamefont {Mathisson}},\ }\href@noop {} {\bibfield  {journal} {\bibinfo  {journal} {Acta Phys. Polon.}\ }\textbf {\bibinfo {volume} {6}},\ \bibinfo {pages} {163} (\bibinfo {year} {1937})}\BibitemShut {NoStop}%
\bibitem [{\citenamefont {Adamczyk}\ \emph {et~al.}(2017)\citenamefont {Adamczyk} \emph {et~al.}}]{STAR:2017ckg}%
  \BibitemOpen
  \bibfield  {author} {\bibinfo {author} {\bibfnamefont {L.}~\bibnamefont {Adamczyk}} \emph {et~al.} (\bibinfo {collaboration} {STAR}),\ }\href {\doibase 10.1038/nature23004} {\bibfield  {journal} {\bibinfo  {journal} {Nature}\ }\textbf {\bibinfo {volume} {548}},\ \bibinfo {pages} {62} (\bibinfo {year} {2017})},\ \Eprint {http://arxiv.org/abs/1701.06657} {arXiv:1701.06657 [nucl-ex]} \BibitemShut {NoStop}%
\bibitem [{\citenamefont {Adam}\ \emph {et~al.}(2018{\natexlab{a}})\citenamefont {Adam} \emph {et~al.}}]{STAR:2018gyt}%
  \BibitemOpen
  \bibfield  {author} {\bibinfo {author} {\bibfnamefont {J.}~\bibnamefont {Adam}} \emph {et~al.} (\bibinfo {collaboration} {STAR}),\ }\href {\doibase 10.1103/PhysRevC.98.014910} {\bibfield  {journal} {\bibinfo  {journal} {Phys. Rev. C}\ }\textbf {\bibinfo {volume} {98}},\ \bibinfo {pages} {014910} (\bibinfo {year} {2018}{\natexlab{a}})},\ \Eprint {http://arxiv.org/abs/1805.04400} {arXiv:1805.04400 [nucl-ex]} \BibitemShut {NoStop}%
\bibitem [{\citenamefont {Acharya}\ \emph {et~al.}(2020)\citenamefont {Acharya} \emph {et~al.}}]{ALICE:2019aid}%
  \BibitemOpen
  \bibfield  {author} {\bibinfo {author} {\bibfnamefont {S.}~\bibnamefont {Acharya}} \emph {et~al.} (\bibinfo {collaboration} {ALICE}),\ }\href {\doibase 10.1103/PhysRevLett.125.012301} {\bibfield  {journal} {\bibinfo  {journal} {Phys. Rev. Lett.}\ }\textbf {\bibinfo {volume} {125}},\ \bibinfo {pages} {012301} (\bibinfo {year} {2020})},\ \Eprint {http://arxiv.org/abs/1910.14408} {arXiv:1910.14408 [nucl-ex]} \BibitemShut {NoStop}%
\bibitem [{\citenamefont {Liang}\ and\ \citenamefont {Wang}(2005{\natexlab{a}})}]{Liang:2004ph}%
  \BibitemOpen
  \bibfield  {author} {\bibinfo {author} {\bibfnamefont {Z.-T.}\ \bibnamefont {Liang}}\ and\ \bibinfo {author} {\bibfnamefont {X.-N.}\ \bibnamefont {Wang}},\ }\href {\doibase 10.1103/PhysRevLett.94.102301} {\bibfield  {journal} {\bibinfo  {journal} {Phys. Rev. Lett.}\ }\textbf {\bibinfo {volume} {94}},\ \bibinfo {pages} {102301} (\bibinfo {year} {2005}{\natexlab{a}})},\ \bibinfo {note} {[Erratum: Phys.Rev.Lett. 96, 039901 (2006)]},\ \Eprint {http://arxiv.org/abs/nucl-th/0410079} {arXiv:nucl-th/0410079} \BibitemShut {NoStop}%
\bibitem [{\citenamefont {Liang}\ and\ \citenamefont {Wang}(2005{\natexlab{b}})}]{Liang:2004xn}%
  \BibitemOpen
  \bibfield  {author} {\bibinfo {author} {\bibfnamefont {Z.-T.}\ \bibnamefont {Liang}}\ and\ \bibinfo {author} {\bibfnamefont {X.-N.}\ \bibnamefont {Wang}},\ }\href {\doibase 10.1016/j.physletb.2005.09.060} {\bibfield  {journal} {\bibinfo  {journal} {Phys. Lett. B}\ }\textbf {\bibinfo {volume} {629}},\ \bibinfo {pages} {20} (\bibinfo {year} {2005}{\natexlab{b}})},\ \Eprint {http://arxiv.org/abs/nucl-th/0411101} {arXiv:nucl-th/0411101} \BibitemShut {NoStop}%
\bibitem [{\citenamefont {Becattini}\ and\ \citenamefont {Tinti}(2010)}]{Becattini:2009wh}%
  \BibitemOpen
  \bibfield  {author} {\bibinfo {author} {\bibfnamefont {F.}~\bibnamefont {Becattini}}\ and\ \bibinfo {author} {\bibfnamefont {L.}~\bibnamefont {Tinti}},\ }\href {\doibase 10.1016/j.aop.2010.03.007} {\bibfield  {journal} {\bibinfo  {journal} {Annals Phys.}\ }\textbf {\bibinfo {volume} {325}},\ \bibinfo {pages} {1566} (\bibinfo {year} {2010})},\ \Eprint {http://arxiv.org/abs/0911.0864} {arXiv:0911.0864 [gr-qc]} \BibitemShut {NoStop}%
\bibitem [{\citenamefont {Becattini}\ and\ \citenamefont {Karpenko}(2018)}]{Becattini:2017gcx}%
  \BibitemOpen
  \bibfield  {author} {\bibinfo {author} {\bibfnamefont {F.}~\bibnamefont {Becattini}}\ and\ \bibinfo {author} {\bibfnamefont {I.}~\bibnamefont {Karpenko}},\ }\href {\doibase 10.1103/PhysRevLett.120.012302} {\bibfield  {journal} {\bibinfo  {journal} {Phys. Rev. Lett.}\ }\textbf {\bibinfo {volume} {120}},\ \bibinfo {pages} {012302} (\bibinfo {year} {2018})},\ \Eprint {http://arxiv.org/abs/1707.07984} {arXiv:1707.07984 [nucl-th]} \BibitemShut {NoStop}%
\bibitem [{\citenamefont {Adam}\ \emph {et~al.}(2018{\natexlab{b}})\citenamefont {Adam} \emph {et~al.}}]{STAR:2018pps}%
  \BibitemOpen
  \bibfield  {author} {\bibinfo {author} {\bibfnamefont {J.}~\bibnamefont {Adam}} \emph {et~al.} (\bibinfo {collaboration} {STAR}),\ }\href {\doibase 10.1103/PhysRevD.98.112009} {\bibfield  {journal} {\bibinfo  {journal} {Phys. Rev. D}\ }\textbf {\bibinfo {volume} {98}},\ \bibinfo {pages} {112009} (\bibinfo {year} {2018}{\natexlab{b}})},\ \Eprint {http://arxiv.org/abs/1808.07634} {arXiv:1808.07634 [hep-ex]} \BibitemShut {NoStop}%
\bibitem [{\citenamefont {Adam}\ \emph {et~al.}(2018{\natexlab{c}})\citenamefont {Adam} \emph {et~al.}}]{STAR:2018fqv}%
  \BibitemOpen
  \bibfield  {author} {\bibinfo {author} {\bibfnamefont {J.}~\bibnamefont {Adam}} \emph {et~al.} (\bibinfo {collaboration} {STAR}),\ }\href {\doibase 10.1103/PhysRevD.98.091103} {\bibfield  {journal} {\bibinfo  {journal} {Phys. Rev. D}\ }\textbf {\bibinfo {volume} {98}},\ \bibinfo {pages} {091103} (\bibinfo {year} {2018}{\natexlab{c}})},\ \Eprint {http://arxiv.org/abs/1808.08000} {arXiv:1808.08000 [hep-ex]} \BibitemShut {NoStop}%
\bibitem [{\citenamefont {Mohanty}\ \emph {et~al.}(2021)\citenamefont {Mohanty}, \citenamefont {Kundu}, \citenamefont {Singha},\ and\ \citenamefont {Singh}}]{Mohanty:2021vbt}%
  \BibitemOpen
  \bibfield  {author} {\bibinfo {author} {\bibfnamefont {B.}~\bibnamefont {Mohanty}}, \bibinfo {author} {\bibfnamefont {S.}~\bibnamefont {Kundu}}, \bibinfo {author} {\bibfnamefont {S.}~\bibnamefont {Singha}}, \ and\ \bibinfo {author} {\bibfnamefont {R.}~\bibnamefont {Singh}},\ }\href {\doibase 10.1142/S0217732321300263} {\bibfield  {journal} {\bibinfo  {journal} {Mod. Phys. Lett. A}\ }\textbf {\bibinfo {volume} {36}},\ \bibinfo {pages} {2130026} (\bibinfo {year} {2021})},\ \Eprint {http://arxiv.org/abs/2112.04816} {arXiv:2112.04816 [nucl-ex]} \BibitemShut {NoStop}%
\bibitem [{\citenamefont {Palermo}\ \emph {et~al.}(2024)\citenamefont {Palermo}, \citenamefont {Grossi}, \citenamefont {Karpenko},\ and\ \citenamefont {Becattini}}]{Palermo:2024tza}%
  \BibitemOpen
  \bibfield  {author} {\bibinfo {author} {\bibfnamefont {A.}~\bibnamefont {Palermo}}, \bibinfo {author} {\bibfnamefont {E.}~\bibnamefont {Grossi}}, \bibinfo {author} {\bibfnamefont {I.}~\bibnamefont {Karpenko}}, \ and\ \bibinfo {author} {\bibfnamefont {F.}~\bibnamefont {Becattini}},\ }\href {\doibase 10.1140/epjc/s10052-024-13229-z} {\bibfield  {journal} {\bibinfo  {journal} {Eur. Phys. J. C}\ }\textbf {\bibinfo {volume} {84}},\ \bibinfo {pages} {920} (\bibinfo {year} {2024})},\ \Eprint {http://arxiv.org/abs/2404.14295} {arXiv:2404.14295 [nucl-th]} \BibitemShut {NoStop}%
\bibitem [{\citenamefont {Becattini}\ \emph {et~al.}(2013)\citenamefont {Becattini}, \citenamefont {Chandra}, \citenamefont {Del~Zanna},\ and\ \citenamefont {Grossi}}]{Becattini:2013fla}%
  \BibitemOpen
  \bibfield  {author} {\bibinfo {author} {\bibfnamefont {F.}~\bibnamefont {Becattini}}, \bibinfo {author} {\bibfnamefont {V.}~\bibnamefont {Chandra}}, \bibinfo {author} {\bibfnamefont {L.}~\bibnamefont {Del~Zanna}}, \ and\ \bibinfo {author} {\bibfnamefont {E.}~\bibnamefont {Grossi}},\ }\href {\doibase 10.1016/j.aop.2013.07.004} {\bibfield  {journal} {\bibinfo  {journal} {Annals Phys.}\ }\textbf {\bibinfo {volume} {338}},\ \bibinfo {pages} {32} (\bibinfo {year} {2013})},\ \Eprint {http://arxiv.org/abs/1303.3431} {arXiv:1303.3431 [nucl-th]} \BibitemShut {NoStop}%
\bibitem [{\citenamefont {Karabali}\ and\ \citenamefont {Nair}(2014)}]{Karabali:2014vla}%
  \BibitemOpen
  \bibfield  {author} {\bibinfo {author} {\bibfnamefont {D.}~\bibnamefont {Karabali}}\ and\ \bibinfo {author} {\bibfnamefont {V.~P.}\ \bibnamefont {Nair}},\ }\href {\doibase 10.1103/PhysRevD.90.105018} {\bibfield  {journal} {\bibinfo  {journal} {Phys. Rev. D}\ }\textbf {\bibinfo {volume} {90}},\ \bibinfo {pages} {105018} (\bibinfo {year} {2014})},\ \Eprint {http://arxiv.org/abs/1406.1551} {arXiv:1406.1551 [hep-th]} \BibitemShut {NoStop}%
\bibitem [{\citenamefont {Florkowski}\ \emph {et~al.}(2018{\natexlab{a}})\citenamefont {Florkowski}, \citenamefont {Friman}, \citenamefont {Jaiswal},\ and\ \citenamefont {Speranza}}]{Florkowski:2017ruc}%
  \BibitemOpen
  \bibfield  {author} {\bibinfo {author} {\bibfnamefont {W.}~\bibnamefont {Florkowski}}, \bibinfo {author} {\bibfnamefont {B.}~\bibnamefont {Friman}}, \bibinfo {author} {\bibfnamefont {A.}~\bibnamefont {Jaiswal}}, \ and\ \bibinfo {author} {\bibfnamefont {E.}~\bibnamefont {Speranza}},\ }\href {\doibase 10.1103/PhysRevC.97.041901} {\bibfield  {journal} {\bibinfo  {journal} {Phys. Rev. C}\ }\textbf {\bibinfo {volume} {97}},\ \bibinfo {pages} {041901} (\bibinfo {year} {2018}{\natexlab{a}})},\ \Eprint {http://arxiv.org/abs/1705.00587} {arXiv:1705.00587 [nucl-th]} \BibitemShut {NoStop}%
\bibitem [{\citenamefont {Florkowski}\ \emph {et~al.}(2018{\natexlab{b}})\citenamefont {Florkowski}, \citenamefont {Friman}, \citenamefont {Jaiswal}, \citenamefont {Ryblewski},\ and\ \citenamefont {Speranza}}]{Florkowski:2017dyn}%
  \BibitemOpen
  \bibfield  {author} {\bibinfo {author} {\bibfnamefont {W.}~\bibnamefont {Florkowski}}, \bibinfo {author} {\bibfnamefont {B.}~\bibnamefont {Friman}}, \bibinfo {author} {\bibfnamefont {A.}~\bibnamefont {Jaiswal}}, \bibinfo {author} {\bibfnamefont {R.}~\bibnamefont {Ryblewski}}, \ and\ \bibinfo {author} {\bibfnamefont {E.}~\bibnamefont {Speranza}},\ }\href {\doibase 10.1103/PhysRevD.97.116017} {\bibfield  {journal} {\bibinfo  {journal} {Phys. Rev. D}\ }\textbf {\bibinfo {volume} {97}},\ \bibinfo {pages} {116017} (\bibinfo {year} {2018}{\natexlab{b}})},\ \Eprint {http://arxiv.org/abs/1712.07676} {arXiv:1712.07676 [nucl-th]} \BibitemShut {NoStop}%
\bibitem [{\citenamefont {Montenegro}\ \emph {et~al.}(2017{\natexlab{a}})\citenamefont {Montenegro}, \citenamefont {Tinti},\ and\ \citenamefont {Torrieri}}]{Montenegro:2017lvf}%
  \BibitemOpen
  \bibfield  {author} {\bibinfo {author} {\bibfnamefont {D.}~\bibnamefont {Montenegro}}, \bibinfo {author} {\bibfnamefont {L.}~\bibnamefont {Tinti}}, \ and\ \bibinfo {author} {\bibfnamefont {G.}~\bibnamefont {Torrieri}},\ }\href {\doibase 10.1103/PhysRevD.96.076016} {\bibfield  {journal} {\bibinfo  {journal} {Phys. Rev. D}\ }\textbf {\bibinfo {volume} {96}},\ \bibinfo {pages} {076016} (\bibinfo {year} {2017}{\natexlab{a}})},\ \Eprint {http://arxiv.org/abs/1703.03079} {arXiv:1703.03079 [hep-th]} \BibitemShut {NoStop}%
\bibitem [{\citenamefont {Montenegro}\ \emph {et~al.}(2017{\natexlab{b}})\citenamefont {Montenegro}, \citenamefont {Tinti},\ and\ \citenamefont {Torrieri}}]{Montenegro:2017rbu}%
  \BibitemOpen
  \bibfield  {author} {\bibinfo {author} {\bibfnamefont {D.}~\bibnamefont {Montenegro}}, \bibinfo {author} {\bibfnamefont {L.}~\bibnamefont {Tinti}}, \ and\ \bibinfo {author} {\bibfnamefont {G.}~\bibnamefont {Torrieri}},\ }\href {\doibase 10.1103/PhysRevD.96.056012} {\bibfield  {journal} {\bibinfo  {journal} {Phys. Rev. D}\ }\textbf {\bibinfo {volume} {96}},\ \bibinfo {pages} {056012} (\bibinfo {year} {2017}{\natexlab{b}})},\ \bibinfo {note} {[Addendum: Phys.Rev.D 96, 079901 (2017)]},\ \Eprint {http://arxiv.org/abs/1701.08263} {arXiv:1701.08263 [hep-th]} \BibitemShut {NoStop}%
\bibitem [{\citenamefont {Montenegro}\ and\ \citenamefont {Torrieri}(2019)}]{Montenegro:2018bcf}%
  \BibitemOpen
  \bibfield  {author} {\bibinfo {author} {\bibfnamefont {D.}~\bibnamefont {Montenegro}}\ and\ \bibinfo {author} {\bibfnamefont {G.}~\bibnamefont {Torrieri}},\ }\href {\doibase 10.1103/PhysRevD.100.056011} {\bibfield  {journal} {\bibinfo  {journal} {Phys. Rev. D}\ }\textbf {\bibinfo {volume} {100}},\ \bibinfo {pages} {056011} (\bibinfo {year} {2019})},\ \Eprint {http://arxiv.org/abs/1807.02796} {arXiv:1807.02796 [hep-th]} \BibitemShut {NoStop}%
\bibitem [{\citenamefont {Becattini}\ \emph {et~al.}(2019)\citenamefont {Becattini}, \citenamefont {Florkowski},\ and\ \citenamefont {Speranza}}]{Becattini:2018duy}%
  \BibitemOpen
  \bibfield  {author} {\bibinfo {author} {\bibfnamefont {F.}~\bibnamefont {Becattini}}, \bibinfo {author} {\bibfnamefont {W.}~\bibnamefont {Florkowski}}, \ and\ \bibinfo {author} {\bibfnamefont {E.}~\bibnamefont {Speranza}},\ }\href {\doibase 10.1016/j.physletb.2018.12.016} {\bibfield  {journal} {\bibinfo  {journal} {Phys. Lett. B}\ }\textbf {\bibinfo {volume} {789}},\ \bibinfo {pages} {419} (\bibinfo {year} {2019})},\ \Eprint {http://arxiv.org/abs/1807.10994} {arXiv:1807.10994 [hep-th]} \BibitemShut {NoStop}%
\bibitem [{\citenamefont {Florkowski}\ \emph {et~al.}(2019{\natexlab{a}})\citenamefont {Florkowski}, \citenamefont {Kumar},\ and\ \citenamefont {Ryblewski}}]{Florkowski:2018fap}%
  \BibitemOpen
  \bibfield  {author} {\bibinfo {author} {\bibfnamefont {W.}~\bibnamefont {Florkowski}}, \bibinfo {author} {\bibfnamefont {A.}~\bibnamefont {Kumar}}, \ and\ \bibinfo {author} {\bibfnamefont {R.}~\bibnamefont {Ryblewski}},\ }\href {\doibase 10.1016/j.ppnp.2019.07.001} {\bibfield  {journal} {\bibinfo  {journal} {Prog. Part. Nucl. Phys.}\ }\textbf {\bibinfo {volume} {108}},\ \bibinfo {pages} {103709} (\bibinfo {year} {2019}{\natexlab{a}})},\ \Eprint {http://arxiv.org/abs/1811.04409} {arXiv:1811.04409 [nucl-th]} \BibitemShut {NoStop}%
\bibitem [{\citenamefont {Florkowski}\ \emph {et~al.}(2019{\natexlab{b}})\citenamefont {Florkowski}, \citenamefont {Kumar}, \citenamefont {Ryblewski},\ and\ \citenamefont {Singh}}]{Florkowski:2019qdp}%
  \BibitemOpen
  \bibfield  {author} {\bibinfo {author} {\bibfnamefont {W.}~\bibnamefont {Florkowski}}, \bibinfo {author} {\bibfnamefont {A.}~\bibnamefont {Kumar}}, \bibinfo {author} {\bibfnamefont {R.}~\bibnamefont {Ryblewski}}, \ and\ \bibinfo {author} {\bibfnamefont {R.}~\bibnamefont {Singh}},\ }\href {\doibase 10.1103/PhysRevC.99.044910} {\bibfield  {journal} {\bibinfo  {journal} {Phys. Rev. C}\ }\textbf {\bibinfo {volume} {99}},\ \bibinfo {pages} {044910} (\bibinfo {year} {2019}{\natexlab{b}})},\ \Eprint {http://arxiv.org/abs/1901.09655} {arXiv:1901.09655 [hep-ph]} \BibitemShut {NoStop}%
\bibitem [{\citenamefont {Montenegro}\ and\ \citenamefont {Torrieri}(2020)}]{Montenegro:2020paq}%
  \BibitemOpen
  \bibfield  {author} {\bibinfo {author} {\bibfnamefont {D.}~\bibnamefont {Montenegro}}\ and\ \bibinfo {author} {\bibfnamefont {G.}~\bibnamefont {Torrieri}},\ }\href {\doibase 10.1103/PhysRevD.102.036007} {\bibfield  {journal} {\bibinfo  {journal} {Phys. Rev. D}\ }\textbf {\bibinfo {volume} {102}},\ \bibinfo {pages} {036007} (\bibinfo {year} {2020})},\ \Eprint {http://arxiv.org/abs/2004.10195} {arXiv:2004.10195 [hep-th]} \BibitemShut {NoStop}%
\bibitem [{\citenamefont {Bhadury}\ \emph {et~al.}(2021{\natexlab{a}})\citenamefont {Bhadury}, \citenamefont {Florkowski}, \citenamefont {Jaiswal}, \citenamefont {Kumar},\ and\ \citenamefont {Ryblewski}}]{Bhadury:2020puc}%
  \BibitemOpen
  \bibfield  {author} {\bibinfo {author} {\bibfnamefont {S.}~\bibnamefont {Bhadury}}, \bibinfo {author} {\bibfnamefont {W.}~\bibnamefont {Florkowski}}, \bibinfo {author} {\bibfnamefont {A.}~\bibnamefont {Jaiswal}}, \bibinfo {author} {\bibfnamefont {A.}~\bibnamefont {Kumar}}, \ and\ \bibinfo {author} {\bibfnamefont {R.}~\bibnamefont {Ryblewski}},\ }\href {\doibase 10.1016/j.physletb.2021.136096} {\bibfield  {journal} {\bibinfo  {journal} {Phys. Lett. B}\ }\textbf {\bibinfo {volume} {814}},\ \bibinfo {pages} {136096} (\bibinfo {year} {2021}{\natexlab{a}})},\ \Eprint {http://arxiv.org/abs/2002.03937} {arXiv:2002.03937 [hep-ph]} \BibitemShut {NoStop}%
\bibitem [{\citenamefont {Becattini}(2021)}]{Becattini:2020sww}%
  \BibitemOpen
  \bibfield  {author} {\bibinfo {author} {\bibfnamefont {F.}~\bibnamefont {Becattini}},\ }\href {\doibase 10.1007/978-3-030-71427-7_2} {\bibfield  {journal} {\bibinfo  {journal} {Lect. Notes Phys.}\ }\textbf {\bibinfo {volume} {987}},\ \bibinfo {pages} {15} (\bibinfo {year} {2021})},\ \Eprint {http://arxiv.org/abs/2004.04050} {arXiv:2004.04050 [hep-th]} \BibitemShut {NoStop}%
\bibitem [{\citenamefont {Shi}\ \emph {et~al.}(2021)\citenamefont {Shi}, \citenamefont {Gale},\ and\ \citenamefont {Jeon}}]{Shi:2020htn}%
  \BibitemOpen
  \bibfield  {author} {\bibinfo {author} {\bibfnamefont {S.}~\bibnamefont {Shi}}, \bibinfo {author} {\bibfnamefont {C.}~\bibnamefont {Gale}}, \ and\ \bibinfo {author} {\bibfnamefont {S.}~\bibnamefont {Jeon}},\ }\href {\doibase 10.1103/PhysRevC.103.044906} {\bibfield  {journal} {\bibinfo  {journal} {Phys. Rev. C}\ }\textbf {\bibinfo {volume} {103}},\ \bibinfo {pages} {044906} (\bibinfo {year} {2021})},\ \Eprint {http://arxiv.org/abs/2008.08618} {arXiv:2008.08618 [nucl-th]} \BibitemShut {NoStop}%
\bibitem [{\citenamefont {Bhadury}\ \emph {et~al.}(2021{\natexlab{b}})\citenamefont {Bhadury}, \citenamefont {Florkowski}, \citenamefont {Jaiswal}, \citenamefont {Kumar},\ and\ \citenamefont {Ryblewski}}]{Bhadury:2020cop}%
  \BibitemOpen
  \bibfield  {author} {\bibinfo {author} {\bibfnamefont {S.}~\bibnamefont {Bhadury}}, \bibinfo {author} {\bibfnamefont {W.}~\bibnamefont {Florkowski}}, \bibinfo {author} {\bibfnamefont {A.}~\bibnamefont {Jaiswal}}, \bibinfo {author} {\bibfnamefont {A.}~\bibnamefont {Kumar}}, \ and\ \bibinfo {author} {\bibfnamefont {R.}~\bibnamefont {Ryblewski}},\ }\href {\doibase 10.1103/PhysRevD.103.014030} {\bibfield  {journal} {\bibinfo  {journal} {Phys. Rev. D}\ }\textbf {\bibinfo {volume} {103}},\ \bibinfo {pages} {014030} (\bibinfo {year} {2021}{\natexlab{b}})},\ \Eprint {http://arxiv.org/abs/2008.10976} {arXiv:2008.10976 [nucl-th]} \BibitemShut {NoStop}%
\bibitem [{\citenamefont {Fu}\ \emph {et~al.}(2021)\citenamefont {Fu}, \citenamefont {Xu}, \citenamefont {Huang},\ and\ \citenamefont {Song}}]{Fu:2020oxj}%
  \BibitemOpen
  \bibfield  {author} {\bibinfo {author} {\bibfnamefont {B.}~\bibnamefont {Fu}}, \bibinfo {author} {\bibfnamefont {K.}~\bibnamefont {Xu}}, \bibinfo {author} {\bibfnamefont {X.-G.}\ \bibnamefont {Huang}}, \ and\ \bibinfo {author} {\bibfnamefont {H.}~\bibnamefont {Song}},\ }\href {\doibase 10.1103/PhysRevC.103.024903} {\bibfield  {journal} {\bibinfo  {journal} {Phys. Rev. C}\ }\textbf {\bibinfo {volume} {103}},\ \bibinfo {pages} {024903} (\bibinfo {year} {2021})},\ \Eprint {http://arxiv.org/abs/2011.03740} {arXiv:2011.03740 [nucl-th]} \BibitemShut {NoStop}%
\bibitem [{\citenamefont {Speranza}\ \emph {et~al.}(2023)\citenamefont {Speranza}, \citenamefont {Bemfica}, \citenamefont {Disconzi},\ and\ \citenamefont {Noronha}}]{Speranza:2021bxf}%
  \BibitemOpen
  \bibfield  {author} {\bibinfo {author} {\bibfnamefont {E.}~\bibnamefont {Speranza}}, \bibinfo {author} {\bibfnamefont {F.~S.}\ \bibnamefont {Bemfica}}, \bibinfo {author} {\bibfnamefont {M.~M.}\ \bibnamefont {Disconzi}}, \ and\ \bibinfo {author} {\bibfnamefont {J.}~\bibnamefont {Noronha}},\ }\href {\doibase 10.1103/PhysRevD.107.054029} {\bibfield  {journal} {\bibinfo  {journal} {Phys. Rev. D}\ }\textbf {\bibinfo {volume} {107}},\ \bibinfo {pages} {054029} (\bibinfo {year} {2023})},\ \Eprint {http://arxiv.org/abs/2104.02110} {arXiv:2104.02110 [hep-th]} \BibitemShut {NoStop}%
\bibitem [{\citenamefont {Wang}\ \emph {et~al.}(2021)\citenamefont {Wang}, \citenamefont {Fang},\ and\ \citenamefont {Pu}}]{Wang:2021ngp}%
  \BibitemOpen
  \bibfield  {author} {\bibinfo {author} {\bibfnamefont {D.-L.}\ \bibnamefont {Wang}}, \bibinfo {author} {\bibfnamefont {S.}~\bibnamefont {Fang}}, \ and\ \bibinfo {author} {\bibfnamefont {S.}~\bibnamefont {Pu}},\ }\href {\doibase 10.1103/PhysRevD.104.114043} {\bibfield  {journal} {\bibinfo  {journal} {Phys. Rev. D}\ }\textbf {\bibinfo {volume} {104}},\ \bibinfo {pages} {114043} (\bibinfo {year} {2021})},\ \Eprint {http://arxiv.org/abs/2107.11726} {arXiv:2107.11726 [nucl-th]} \BibitemShut {NoStop}%
\bibitem [{\citenamefont {Hongo}\ \emph {et~al.}(2021)\citenamefont {Hongo}, \citenamefont {Huang}, \citenamefont {Kaminski}, \citenamefont {Stephanov},\ and\ \citenamefont {Yee}}]{Hongo:2021ona}%
  \BibitemOpen
  \bibfield  {author} {\bibinfo {author} {\bibfnamefont {M.}~\bibnamefont {Hongo}}, \bibinfo {author} {\bibfnamefont {X.-G.}\ \bibnamefont {Huang}}, \bibinfo {author} {\bibfnamefont {M.}~\bibnamefont {Kaminski}}, \bibinfo {author} {\bibfnamefont {M.}~\bibnamefont {Stephanov}}, \ and\ \bibinfo {author} {\bibfnamefont {H.-U.}\ \bibnamefont {Yee}},\ }\href {\doibase 10.1007/JHEP11(2021)150} {\bibfield  {journal} {\bibinfo  {journal} {JHEP}\ }\textbf {\bibinfo {volume} {11}},\ \bibinfo {pages} {150} (\bibinfo {year} {2021})},\ \Eprint {http://arxiv.org/abs/2107.14231} {arXiv:2107.14231 [hep-th]} \BibitemShut {NoStop}%
\bibitem [{\citenamefont {Hu}(2022{\natexlab{a}})}]{Hu:2021pwh}%
  \BibitemOpen
  \bibfield  {author} {\bibinfo {author} {\bibfnamefont {J.}~\bibnamefont {Hu}},\ }\href {\doibase 10.1103/PhysRevD.105.076009} {\bibfield  {journal} {\bibinfo  {journal} {Phys. Rev. D}\ }\textbf {\bibinfo {volume} {105}},\ \bibinfo {pages} {076009} (\bibinfo {year} {2022}{\natexlab{a}})},\ \Eprint {http://arxiv.org/abs/2111.03571} {arXiv:2111.03571 [hep-ph]} \BibitemShut {NoStop}%
\bibitem [{\citenamefont {Wang}\ \emph {et~al.}(2022)\citenamefont {Wang}, \citenamefont {Xie}, \citenamefont {Fang},\ and\ \citenamefont {Pu}}]{Wang:2021wqq}%
  \BibitemOpen
  \bibfield  {author} {\bibinfo {author} {\bibfnamefont {D.-L.}\ \bibnamefont {Wang}}, \bibinfo {author} {\bibfnamefont {X.-Q.}\ \bibnamefont {Xie}}, \bibinfo {author} {\bibfnamefont {S.}~\bibnamefont {Fang}}, \ and\ \bibinfo {author} {\bibfnamefont {S.}~\bibnamefont {Pu}},\ }\href {\doibase 10.1103/PhysRevD.105.114050} {\bibfield  {journal} {\bibinfo  {journal} {Phys. Rev. D}\ }\textbf {\bibinfo {volume} {105}},\ \bibinfo {pages} {114050} (\bibinfo {year} {2022})},\ \Eprint {http://arxiv.org/abs/2112.15535} {arXiv:2112.15535 [hep-ph]} \BibitemShut {NoStop}%
\bibitem [{\citenamefont {Weickgenannt}\ \emph {et~al.}(2022)\citenamefont {Weickgenannt}, \citenamefont {Wagner}, \citenamefont {Speranza},\ and\ \citenamefont {Rischke}}]{Weickgenannt:2022zxs}%
  \BibitemOpen
  \bibfield  {author} {\bibinfo {author} {\bibfnamefont {N.}~\bibnamefont {Weickgenannt}}, \bibinfo {author} {\bibfnamefont {D.}~\bibnamefont {Wagner}}, \bibinfo {author} {\bibfnamefont {E.}~\bibnamefont {Speranza}}, \ and\ \bibinfo {author} {\bibfnamefont {D.~H.}\ \bibnamefont {Rischke}},\ }\href {\doibase 10.1103/PhysRevD.106.096014} {\bibfield  {journal} {\bibinfo  {journal} {Phys. Rev. D}\ }\textbf {\bibinfo {volume} {106}},\ \bibinfo {pages} {096014} (\bibinfo {year} {2022})},\ \Eprint {http://arxiv.org/abs/2203.04766} {arXiv:2203.04766 [nucl-th]} \BibitemShut {NoStop}%
\bibitem [{\citenamefont {Gallegos}\ \emph {et~al.}(2023)\citenamefont {Gallegos}, \citenamefont {Gursoy},\ and\ \citenamefont {Yarom}}]{Gallegos:2022jow}%
  \BibitemOpen
  \bibfield  {author} {\bibinfo {author} {\bibfnamefont {A.~D.}\ \bibnamefont {Gallegos}}, \bibinfo {author} {\bibfnamefont {U.}~\bibnamefont {Gursoy}}, \ and\ \bibinfo {author} {\bibfnamefont {A.}~\bibnamefont {Yarom}},\ }\href {\doibase 10.1007/JHEP05(2023)139} {\bibfield  {journal} {\bibinfo  {journal} {JHEP}\ }\textbf {\bibinfo {volume} {05}},\ \bibinfo {pages} {139} (\bibinfo {year} {2023})},\ \Eprint {http://arxiv.org/abs/2203.05044} {arXiv:2203.05044 [hep-th]} \BibitemShut {NoStop}%
\bibitem [{\citenamefont {Sarwar}\ \emph {et~al.}(2023)\citenamefont {Sarwar}, \citenamefont {Hasanujjaman}, \citenamefont {Bhatt}, \citenamefont {Mishra},\ and\ \citenamefont {Alam}}]{Sarwar:2022yzs}%
  \BibitemOpen
  \bibfield  {author} {\bibinfo {author} {\bibfnamefont {G.}~\bibnamefont {Sarwar}}, \bibinfo {author} {\bibfnamefont {M.}~\bibnamefont {Hasanujjaman}}, \bibinfo {author} {\bibfnamefont {J.~R.}\ \bibnamefont {Bhatt}}, \bibinfo {author} {\bibfnamefont {H.}~\bibnamefont {Mishra}}, \ and\ \bibinfo {author} {\bibfnamefont {J.-e.}\ \bibnamefont {Alam}},\ }\href {\doibase 10.1103/PhysRevD.107.054031} {\bibfield  {journal} {\bibinfo  {journal} {Phys. Rev. D}\ }\textbf {\bibinfo {volume} {107}},\ \bibinfo {pages} {054031} (\bibinfo {year} {2023})},\ \Eprint {http://arxiv.org/abs/2209.08652} {arXiv:2209.08652 [nucl-th]} \BibitemShut {NoStop}%
\bibitem [{\citenamefont {Biswas}\ \emph {et~al.}(2023{\natexlab{a}})\citenamefont {Biswas}, \citenamefont {Daher}, \citenamefont {Das}, \citenamefont {Florkowski},\ and\ \citenamefont {Ryblewski}}]{Biswas:2022bht}%
  \BibitemOpen
  \bibfield  {author} {\bibinfo {author} {\bibfnamefont {R.}~\bibnamefont {Biswas}}, \bibinfo {author} {\bibfnamefont {A.}~\bibnamefont {Daher}}, \bibinfo {author} {\bibfnamefont {A.}~\bibnamefont {Das}}, \bibinfo {author} {\bibfnamefont {W.}~\bibnamefont {Florkowski}}, \ and\ \bibinfo {author} {\bibfnamefont {R.}~\bibnamefont {Ryblewski}},\ }\href {\doibase 10.1103/PhysRevD.107.094022} {\bibfield  {journal} {\bibinfo  {journal} {Phys. Rev. D}\ }\textbf {\bibinfo {volume} {107}},\ \bibinfo {pages} {094022} (\bibinfo {year} {2023}{\natexlab{a}})},\ \Eprint {http://arxiv.org/abs/2211.02934} {arXiv:2211.02934 [nucl-th]} \BibitemShut {NoStop}%
\bibitem [{\citenamefont {Biswas}\ \emph {et~al.}(2023{\natexlab{b}})\citenamefont {Biswas}, \citenamefont {Daher}, \citenamefont {Das}, \citenamefont {Florkowski},\ and\ \citenamefont {Ryblewski}}]{Biswas:2023qsw}%
  \BibitemOpen
  \bibfield  {author} {\bibinfo {author} {\bibfnamefont {R.}~\bibnamefont {Biswas}}, \bibinfo {author} {\bibfnamefont {A.}~\bibnamefont {Daher}}, \bibinfo {author} {\bibfnamefont {A.}~\bibnamefont {Das}}, \bibinfo {author} {\bibfnamefont {W.}~\bibnamefont {Florkowski}}, \ and\ \bibinfo {author} {\bibfnamefont {R.}~\bibnamefont {Ryblewski}},\ }\href {\doibase 10.1103/PhysRevD.108.014024} {\bibfield  {journal} {\bibinfo  {journal} {Phys. Rev. D}\ }\textbf {\bibinfo {volume} {108}},\ \bibinfo {pages} {014024} (\bibinfo {year} {2023}{\natexlab{b}})},\ \Eprint {http://arxiv.org/abs/2304.01009} {arXiv:2304.01009 [nucl-th]} \BibitemShut {NoStop}%
\bibitem [{\citenamefont {Abboud}\ \emph {et~al.}(2024)\citenamefont {Abboud}, \citenamefont {Speranza},\ and\ \citenamefont {Noronha}}]{Abboud:2023hos}%
  \BibitemOpen
  \bibfield  {author} {\bibinfo {author} {\bibfnamefont {N.}~\bibnamefont {Abboud}}, \bibinfo {author} {\bibfnamefont {E.}~\bibnamefont {Speranza}}, \ and\ \bibinfo {author} {\bibfnamefont {J.}~\bibnamefont {Noronha}},\ }\href {\doibase 10.1103/PhysRevD.109.094007} {\bibfield  {journal} {\bibinfo  {journal} {Phys. Rev. D}\ }\textbf {\bibinfo {volume} {109}},\ \bibinfo {pages} {094007} (\bibinfo {year} {2024})},\ \Eprint {http://arxiv.org/abs/2308.02928} {arXiv:2308.02928 [hep-th]} \BibitemShut {NoStop}%
\bibitem [{\citenamefont {Kiamari}\ \emph {et~al.}(2024)\citenamefont {Kiamari}, \citenamefont {Sadooghi},\ and\ \citenamefont {Jafari}}]{Kiamari:2023fbe}%
  \BibitemOpen
  \bibfield  {author} {\bibinfo {author} {\bibfnamefont {M.}~\bibnamefont {Kiamari}}, \bibinfo {author} {\bibfnamefont {N.}~\bibnamefont {Sadooghi}}, \ and\ \bibinfo {author} {\bibfnamefont {M.~S.}\ \bibnamefont {Jafari}},\ }\href {\doibase 10.1103/PhysRevD.109.036024} {\bibfield  {journal} {\bibinfo  {journal} {Phys. Rev. D}\ }\textbf {\bibinfo {volume} {109}},\ \bibinfo {pages} {036024} (\bibinfo {year} {2024})},\ \Eprint {http://arxiv.org/abs/2310.01874} {arXiv:2310.01874 [nucl-th]} \BibitemShut {NoStop}%
\bibitem [{\citenamefont {Weickgenannt}\ and\ \citenamefont {Blaizot}(2024)}]{Weickgenannt:2023nge}%
  \BibitemOpen
  \bibfield  {author} {\bibinfo {author} {\bibfnamefont {N.}~\bibnamefont {Weickgenannt}}\ and\ \bibinfo {author} {\bibfnamefont {J.-P.}\ \bibnamefont {Blaizot}},\ }\href {\doibase 10.1103/PhysRevD.109.056012} {\bibfield  {journal} {\bibinfo  {journal} {Phys. Rev. D}\ }\textbf {\bibinfo {volume} {109}},\ \bibinfo {pages} {056012} (\bibinfo {year} {2024})},\ \Eprint {http://arxiv.org/abs/2311.15817} {arXiv:2311.15817 [hep-ph]} \BibitemShut {NoStop}%
\bibitem [{\citenamefont {Ren}\ \emph {et~al.}(2024)\citenamefont {Ren}, \citenamefont {Yang}, \citenamefont {Wang},\ and\ \citenamefont {Pu}}]{Ren:2024pur}%
  \BibitemOpen
  \bibfield  {author} {\bibinfo {author} {\bibfnamefont {X.}~\bibnamefont {Ren}}, \bibinfo {author} {\bibfnamefont {C.}~\bibnamefont {Yang}}, \bibinfo {author} {\bibfnamefont {D.-L.}\ \bibnamefont {Wang}}, \ and\ \bibinfo {author} {\bibfnamefont {S.}~\bibnamefont {Pu}},\ }\href {\doibase 10.1103/PhysRevD.110.034010} {\bibfield  {journal} {\bibinfo  {journal} {Phys. Rev. D}\ }\textbf {\bibinfo {volume} {110}},\ \bibinfo {pages} {034010} (\bibinfo {year} {2024})},\ \Eprint {http://arxiv.org/abs/2405.03105} {arXiv:2405.03105 [nucl-th]} \BibitemShut {NoStop}%
\bibitem [{\citenamefont {Florkowski}\ and\ \citenamefont {Hontarenko}(2025)}]{Florkowski:2024bfw}%
  \BibitemOpen
  \bibfield  {author} {\bibinfo {author} {\bibfnamefont {W.}~\bibnamefont {Florkowski}}\ and\ \bibinfo {author} {\bibfnamefont {M.}~\bibnamefont {Hontarenko}},\ }\href {\doibase 10.1103/PhysRevLett.134.082302} {\bibfield  {journal} {\bibinfo  {journal} {Phys. Rev. Lett.}\ }\textbf {\bibinfo {volume} {134}},\ \bibinfo {pages} {082302} (\bibinfo {year} {2025})},\ \Eprint {http://arxiv.org/abs/2405.03263} {arXiv:2405.03263 [hep-ph]} \BibitemShut {NoStop}%
\bibitem [{\citenamefont {Lin}\ and\ \citenamefont {Tang}(2024)}]{Lin:2024cxo}%
  \BibitemOpen
  \bibfield  {author} {\bibinfo {author} {\bibfnamefont {S.}~\bibnamefont {Lin}}\ and\ \bibinfo {author} {\bibfnamefont {H.}~\bibnamefont {Tang}},\ }\href {\doibase 10.1103/PhysRevD.110.074042} {\bibfield  {journal} {\bibinfo  {journal} {Phys. Rev. D}\ }\textbf {\bibinfo {volume} {110}},\ \bibinfo {pages} {074042} (\bibinfo {year} {2024})},\ \Eprint {http://arxiv.org/abs/2406.17632} {arXiv:2406.17632 [nucl-th]} \BibitemShut {NoStop}%
\bibitem [{\citenamefont {Buzzegoli}\ and\ \citenamefont {Palermo}(2024)}]{Buzzegoli:2024mra}%
  \BibitemOpen
  \bibfield  {author} {\bibinfo {author} {\bibfnamefont {M.}~\bibnamefont {Buzzegoli}}\ and\ \bibinfo {author} {\bibfnamefont {A.}~\bibnamefont {Palermo}},\ }\href {\doibase 10.1103/PhysRevLett.133.262301} {\bibfield  {journal} {\bibinfo  {journal} {Phys. Rev. Lett.}\ }\textbf {\bibinfo {volume} {133}},\ \bibinfo {pages} {262301} (\bibinfo {year} {2024})},\ \Eprint {http://arxiv.org/abs/2407.14345} {arXiv:2407.14345 [hep-ph]} \BibitemShut {NoStop}%
\bibitem [{\citenamefont {Fang}\ and\ \citenamefont {Pu}(2025)}]{Fang:2024vds}%
  \BibitemOpen
  \bibfield  {author} {\bibinfo {author} {\bibfnamefont {S.}~\bibnamefont {Fang}}\ and\ \bibinfo {author} {\bibfnamefont {S.}~\bibnamefont {Pu}},\ }\href {\doibase 10.1103/PhysRevD.111.034015} {\bibfield  {journal} {\bibinfo  {journal} {Phys. Rev. D}\ }\textbf {\bibinfo {volume} {111}},\ \bibinfo {pages} {034015} (\bibinfo {year} {2025})},\ \Eprint {http://arxiv.org/abs/2408.09877} {arXiv:2408.09877 [hep-ph]} \BibitemShut {NoStop}%
\bibitem [{\citenamefont {Tiwari}\ and\ \citenamefont {Patra}(2024)}]{Tiwari:2024trl}%
  \BibitemOpen
  \bibfield  {author} {\bibinfo {author} {\bibfnamefont {A.}~\bibnamefont {Tiwari}}\ and\ \bibinfo {author} {\bibfnamefont {B.~K.}\ \bibnamefont {Patra}},\ }\href@noop {} {\  (\bibinfo {year} {2024})},\ \Eprint {http://arxiv.org/abs/2408.11514} {arXiv:2408.11514 [hep-th]} \BibitemShut {NoStop}%
\bibitem [{\citenamefont {Fang}\ \emph {et~al.}(2024)\citenamefont {Fang}, \citenamefont {Hattori},\ and\ \citenamefont {Hu}}]{Fang:2024sym}%
  \BibitemOpen
  \bibfield  {author} {\bibinfo {author} {\bibfnamefont {Z.}~\bibnamefont {Fang}}, \bibinfo {author} {\bibfnamefont {K.}~\bibnamefont {Hattori}}, \ and\ \bibinfo {author} {\bibfnamefont {J.}~\bibnamefont {Hu}},\ }\href@noop {} {\  (\bibinfo {year} {2024})},\ \Eprint {http://arxiv.org/abs/2409.07096} {arXiv:2409.07096 [hep-ph]} \BibitemShut {NoStop}%
\bibitem [{\citenamefont {Wagner}(2025)}]{Wagner:2024fry}%
  \BibitemOpen
  \bibfield  {author} {\bibinfo {author} {\bibfnamefont {D.}~\bibnamefont {Wagner}},\ }\href {\doibase 10.1103/PhysRevD.111.016008} {\bibfield  {journal} {\bibinfo  {journal} {Phys. Rev. D}\ }\textbf {\bibinfo {volume} {111}},\ \bibinfo {pages} {016008} (\bibinfo {year} {2025})},\ \Eprint {http://arxiv.org/abs/2409.07143} {arXiv:2409.07143 [nucl-th]} \BibitemShut {NoStop}%
\bibitem [{\citenamefont {Weickgenannt}\ and\ \citenamefont {Blaizot}(2025)}]{Weickgenannt:2024ibf}%
  \BibitemOpen
  \bibfield  {author} {\bibinfo {author} {\bibfnamefont {N.}~\bibnamefont {Weickgenannt}}\ and\ \bibinfo {author} {\bibfnamefont {J.-P.}\ \bibnamefont {Blaizot}},\ }\href {\doibase 10.1103/PhysRevD.111.056006} {\bibfield  {journal} {\bibinfo  {journal} {Phys. Rev. D}\ }\textbf {\bibinfo {volume} {111}},\ \bibinfo {pages} {056006} (\bibinfo {year} {2025})},\ \Eprint {http://arxiv.org/abs/2409.11045} {arXiv:2409.11045 [hep-ph]} \BibitemShut {NoStop}%
\bibitem [{\citenamefont {Dey}\ and\ \citenamefont {Das}(2024)}]{Dey:2024cwo}%
  \BibitemOpen
  \bibfield  {author} {\bibinfo {author} {\bibfnamefont {S.}~\bibnamefont {Dey}}\ and\ \bibinfo {author} {\bibfnamefont {A.}~\bibnamefont {Das}},\ }\href@noop {} {\  (\bibinfo {year} {2024})},\ \Eprint {http://arxiv.org/abs/2410.04141} {arXiv:2410.04141 [nucl-th]} \BibitemShut {NoStop}%
\bibitem [{\citenamefont {She}\ \emph {et~al.}(2025)\citenamefont {She}, \citenamefont {Qiu},\ and\ \citenamefont {Hou}}]{She:2024rnx}%
  \BibitemOpen
  \bibfield  {author} {\bibinfo {author} {\bibfnamefont {D.}~\bibnamefont {She}}, \bibinfo {author} {\bibfnamefont {Y.-W.}\ \bibnamefont {Qiu}}, \ and\ \bibinfo {author} {\bibfnamefont {D.}~\bibnamefont {Hou}},\ }\href {\doibase 10.1103/PhysRevD.111.036027} {\bibfield  {journal} {\bibinfo  {journal} {Phys. Rev. D}\ }\textbf {\bibinfo {volume} {111}},\ \bibinfo {pages} {036027} (\bibinfo {year} {2025})},\ \Eprint {http://arxiv.org/abs/2410.15142} {arXiv:2410.15142 [nucl-th]} \BibitemShut {NoStop}%
\bibitem [{\citenamefont {Singh}\ \emph {et~al.}(2025)\citenamefont {Singh}, \citenamefont {Ryblewski},\ and\ \citenamefont {Florkowski}}]{Singh:2024cub}%
  \BibitemOpen
  \bibfield  {author} {\bibinfo {author} {\bibfnamefont {S.~K.}\ \bibnamefont {Singh}}, \bibinfo {author} {\bibfnamefont {R.}~\bibnamefont {Ryblewski}}, \ and\ \bibinfo {author} {\bibfnamefont {W.}~\bibnamefont {Florkowski}},\ }\href {\doibase 10.1103/PhysRevC.111.024907} {\bibfield  {journal} {\bibinfo  {journal} {Phys. Rev. C}\ }\textbf {\bibinfo {volume} {111}},\ \bibinfo {pages} {024907} (\bibinfo {year} {2025})},\ \Eprint {http://arxiv.org/abs/2411.08223} {arXiv:2411.08223 [hep-ph]} \BibitemShut {NoStop}%
\bibitem [{\citenamefont {Huang}(2024)}]{Huang:2024ffg}%
  \BibitemOpen
  \bibfield  {author} {\bibinfo {author} {\bibfnamefont {X.-G.}\ \bibnamefont {Huang}},\ }\href@noop {} {\  (\bibinfo {year} {2024})},\ \Eprint {http://arxiv.org/abs/2411.11753} {arXiv:2411.11753 [nucl-th]} \BibitemShut {NoStop}%
\bibitem [{\citenamefont {Daher}\ \emph {et~al.}(2025)\citenamefont {Daher}, \citenamefont {Sheng}, \citenamefont {Wagner},\ and\ \citenamefont {Becattini}}]{Daher:2025pfq}%
  \BibitemOpen
  \bibfield  {author} {\bibinfo {author} {\bibfnamefont {A.}~\bibnamefont {Daher}}, \bibinfo {author} {\bibfnamefont {X.-L.}\ \bibnamefont {Sheng}}, \bibinfo {author} {\bibfnamefont {D.}~\bibnamefont {Wagner}}, \ and\ \bibinfo {author} {\bibfnamefont {F.}~\bibnamefont {Becattini}},\ }\href@noop {} {\  (\bibinfo {year} {2025})},\ \Eprint {http://arxiv.org/abs/2503.03713} {arXiv:2503.03713 [nucl-th]} \BibitemShut {NoStop}%
\bibitem [{\citenamefont {Anderson}\ and\ \citenamefont {Witting}(1974)}]{anderson1974relativistic}%
  \BibitemOpen
  \bibfield  {author} {\bibinfo {author} {\bibfnamefont {J.~L.}\ \bibnamefont {Anderson}}\ and\ \bibinfo {author} {\bibfnamefont {H.}~\bibnamefont {Witting}},\ }\href@noop {} {\bibfield  {journal} {\bibinfo  {journal} {Physica}\ }\textbf {\bibinfo {volume} {74}},\ \bibinfo {pages} {466} (\bibinfo {year} {1974})}\BibitemShut {NoStop}%
\bibitem [{\citenamefont {Jaiswal}\ \emph {et~al.}(2013)\citenamefont {Jaiswal}, \citenamefont {Bhalerao},\ and\ \citenamefont {Pal}}]{Jaiswal:2012qm}%
  \BibitemOpen
  \bibfield  {author} {\bibinfo {author} {\bibfnamefont {A.}~\bibnamefont {Jaiswal}}, \bibinfo {author} {\bibfnamefont {R.~S.}\ \bibnamefont {Bhalerao}}, \ and\ \bibinfo {author} {\bibfnamefont {S.}~\bibnamefont {Pal}},\ }\href {\doibase 10.1016/j.physletb.2013.02.025} {\bibfield  {journal} {\bibinfo  {journal} {Phys. Lett. B}\ }\textbf {\bibinfo {volume} {720}},\ \bibinfo {pages} {347} (\bibinfo {year} {2013})},\ \Eprint {http://arxiv.org/abs/1204.3779} {arXiv:1204.3779 [nucl-th]} \BibitemShut {NoStop}%
\bibitem [{\citenamefont {Wagner}\ \emph {et~al.}(2023)\citenamefont {Wagner}, \citenamefont {Weickgenannt},\ and\ \citenamefont {Speranza}}]{Wagner:2023cct}%
  \BibitemOpen
  \bibfield  {author} {\bibinfo {author} {\bibfnamefont {D.}~\bibnamefont {Wagner}}, \bibinfo {author} {\bibfnamefont {N.}~\bibnamefont {Weickgenannt}}, \ and\ \bibinfo {author} {\bibfnamefont {E.}~\bibnamefont {Speranza}},\ }\href {\doibase 10.1103/PhysRevD.108.116017} {\bibfield  {journal} {\bibinfo  {journal} {Phys. Rev. D}\ }\textbf {\bibinfo {volume} {108}},\ \bibinfo {pages} {116017} (\bibinfo {year} {2023})},\ \Eprint {http://arxiv.org/abs/2306.05936} {arXiv:2306.05936 [nucl-th]} \BibitemShut {NoStop}%
\bibitem [{\citenamefont {Kumar}\ \emph {et~al.}(2024)\citenamefont {Kumar}, \citenamefont {Yang},\ and\ \citenamefont {Gubler}}]{Kumar:2023ojl}%
  \BibitemOpen
  \bibfield  {author} {\bibinfo {author} {\bibfnamefont {A.}~\bibnamefont {Kumar}}, \bibinfo {author} {\bibfnamefont {D.-L.}\ \bibnamefont {Yang}}, \ and\ \bibinfo {author} {\bibfnamefont {P.}~\bibnamefont {Gubler}},\ }\href {\doibase 10.1103/PhysRevD.109.054038} {\bibfield  {journal} {\bibinfo  {journal} {Phys. Rev. D}\ }\textbf {\bibinfo {volume} {109}},\ \bibinfo {pages} {054038} (\bibinfo {year} {2024})},\ \Eprint {http://arxiv.org/abs/2312.16900} {arXiv:2312.16900 [nucl-th]} \BibitemShut {NoStop}%
\bibitem [{\citenamefont {Weickgenannt}\ \emph {et~al.}(2021{\natexlab{a}})\citenamefont {Weickgenannt}, \citenamefont {Speranza}, \citenamefont {Sheng}, \citenamefont {Wang},\ and\ \citenamefont {Rischke}}]{Weickgenannt:2021cuo}%
  \BibitemOpen
  \bibfield  {author} {\bibinfo {author} {\bibfnamefont {N.}~\bibnamefont {Weickgenannt}}, \bibinfo {author} {\bibfnamefont {E.}~\bibnamefont {Speranza}}, \bibinfo {author} {\bibfnamefont {X.-l.}\ \bibnamefont {Sheng}}, \bibinfo {author} {\bibfnamefont {Q.}~\bibnamefont {Wang}}, \ and\ \bibinfo {author} {\bibfnamefont {D.~H.}\ \bibnamefont {Rischke}},\ }\href {\doibase 10.1103/PhysRevD.104.016022} {\bibfield  {journal} {\bibinfo  {journal} {Phys. Rev. D}\ }\textbf {\bibinfo {volume} {104}},\ \bibinfo {pages} {016022} (\bibinfo {year} {2021}{\natexlab{a}})},\ \Eprint {http://arxiv.org/abs/2103.04896} {arXiv:2103.04896 [nucl-th]} \BibitemShut {NoStop}%
\bibitem [{\citenamefont {Bhadury}\ \emph {et~al.}(2022)\citenamefont {Bhadury}, \citenamefont {Florkowski}, \citenamefont {Jaiswal}, \citenamefont {Kumar},\ and\ \citenamefont {Ryblewski}}]{Bhadury:2022ulr}%
  \BibitemOpen
  \bibfield  {author} {\bibinfo {author} {\bibfnamefont {S.}~\bibnamefont {Bhadury}}, \bibinfo {author} {\bibfnamefont {W.}~\bibnamefont {Florkowski}}, \bibinfo {author} {\bibfnamefont {A.}~\bibnamefont {Jaiswal}}, \bibinfo {author} {\bibfnamefont {A.}~\bibnamefont {Kumar}}, \ and\ \bibinfo {author} {\bibfnamefont {R.}~\bibnamefont {Ryblewski}},\ }\href {\doibase 10.1103/PhysRevLett.129.192301} {\bibfield  {journal} {\bibinfo  {journal} {Phys. Rev. Lett.}\ }\textbf {\bibinfo {volume} {129}},\ \bibinfo {pages} {192301} (\bibinfo {year} {2022})},\ \Eprint {http://arxiv.org/abs/2204.01357} {arXiv:2204.01357 [nucl-th]} \BibitemShut {NoStop}%
\bibitem [{\citenamefont {Bhadury}(2024)}]{Bhadury:2024ckc}%
  \BibitemOpen
  \bibfield  {author} {\bibinfo {author} {\bibfnamefont {S.}~\bibnamefont {Bhadury}},\ }\href@noop {} {\  (\bibinfo {year} {2024})},\ \Eprint {http://arxiv.org/abs/2408.14462} {arXiv:2408.14462 [hep-ph]} \BibitemShut {NoStop}%
\bibitem [{\citenamefont {Landau}\ and\ \citenamefont {Lifshitz}(1987)}]{landau1987fluid}%
  \BibitemOpen
  \bibfield  {author} {\bibinfo {author} {\bibfnamefont {L.}~\bibnamefont {Landau}}\ and\ \bibinfo {author} {\bibfnamefont {E.}~\bibnamefont {Lifshitz}},\ }\href@noop {} {\emph {\bibinfo {title} {Fluid mechanics.}}},\ Vol.~\bibinfo {volume} {6}\ (\bibinfo  {publisher} {Elsevier},\ \bibinfo {year} {1987})\BibitemShut {NoStop}%
\bibitem [{\citenamefont {Teaney}\ and\ \citenamefont {Yan}(2014)}]{Teaney:2013gca}%
  \BibitemOpen
  \bibfield  {author} {\bibinfo {author} {\bibfnamefont {D.}~\bibnamefont {Teaney}}\ and\ \bibinfo {author} {\bibfnamefont {L.}~\bibnamefont {Yan}},\ }\href {\doibase 10.1103/PhysRevC.89.014901} {\bibfield  {journal} {\bibinfo  {journal} {Phys. Rev. C}\ }\textbf {\bibinfo {volume} {89}},\ \bibinfo {pages} {014901} (\bibinfo {year} {2014})},\ \Eprint {http://arxiv.org/abs/1304.3753} {arXiv:1304.3753 [nucl-th]} \BibitemShut {NoStop}%
\bibitem [{\citenamefont {Dash}\ \emph {et~al.}(2022)\citenamefont {Dash}, \citenamefont {Bhadury}, \citenamefont {Jaiswal},\ and\ \citenamefont {Jaiswal}}]{Dash:2021ibx}%
  \BibitemOpen
  \bibfield  {author} {\bibinfo {author} {\bibfnamefont {D.}~\bibnamefont {Dash}}, \bibinfo {author} {\bibfnamefont {S.}~\bibnamefont {Bhadury}}, \bibinfo {author} {\bibfnamefont {S.}~\bibnamefont {Jaiswal}}, \ and\ \bibinfo {author} {\bibfnamefont {A.}~\bibnamefont {Jaiswal}},\ }\href {\doibase 10.1016/j.physletb.2022.137202} {\bibfield  {journal} {\bibinfo  {journal} {Phys. Lett. B}\ }\textbf {\bibinfo {volume} {831}},\ \bibinfo {pages} {137202} (\bibinfo {year} {2022})},\ \Eprint {http://arxiv.org/abs/2112.14581} {arXiv:2112.14581 [nucl-th]} \BibitemShut {NoStop}%
\bibitem [{\citenamefont {Kamata}(2024)}]{Kamata:2022ola}%
  \BibitemOpen
  \bibfield  {author} {\bibinfo {author} {\bibfnamefont {S.}~\bibnamefont {Kamata}},\ }\href {\doibase 10.1103/PhysRevD.110.116020} {\bibfield  {journal} {\bibinfo  {journal} {Phys. Rev. D}\ }\textbf {\bibinfo {volume} {110}},\ \bibinfo {pages} {116020} (\bibinfo {year} {2024})},\ \Eprint {http://arxiv.org/abs/2212.14506} {arXiv:2212.14506 [hep-th]} \BibitemShut {NoStop}%
\bibitem [{\citenamefont {Singh}\ \emph {et~al.}(2024)\citenamefont {Singh}, \citenamefont {Kurian},\ and\ \citenamefont {Chandra}}]{Singh:2024leo}%
  \BibitemOpen
  \bibfield  {author} {\bibinfo {author} {\bibfnamefont {S.~K.}\ \bibnamefont {Singh}}, \bibinfo {author} {\bibfnamefont {M.}~\bibnamefont {Kurian}}, \ and\ \bibinfo {author} {\bibfnamefont {V.}~\bibnamefont {Chandra}},\ }\href {\doibase 10.1103/PhysRevD.110.014004} {\bibfield  {journal} {\bibinfo  {journal} {Phys. Rev. D}\ }\textbf {\bibinfo {volume} {110}},\ \bibinfo {pages} {014004} (\bibinfo {year} {2024})},\ \Eprint {http://arxiv.org/abs/2403.13160} {arXiv:2403.13160 [hep-ph]} \BibitemShut {NoStop}%
\bibitem [{\citenamefont {Bemfica}\ \emph {et~al.}(2018)\citenamefont {Bemfica}, \citenamefont {Disconzi},\ and\ \citenamefont {Noronha}}]{Bemfica:2017wps}%
  \BibitemOpen
  \bibfield  {author} {\bibinfo {author} {\bibfnamefont {F.~S.}\ \bibnamefont {Bemfica}}, \bibinfo {author} {\bibfnamefont {M.~M.}\ \bibnamefont {Disconzi}}, \ and\ \bibinfo {author} {\bibfnamefont {J.}~\bibnamefont {Noronha}},\ }\href {\doibase 10.1103/PhysRevD.98.104064} {\bibfield  {journal} {\bibinfo  {journal} {Phys. Rev. D}\ }\textbf {\bibinfo {volume} {98}},\ \bibinfo {pages} {104064} (\bibinfo {year} {2018})},\ \Eprint {http://arxiv.org/abs/1708.06255} {arXiv:1708.06255 [gr-qc]} \BibitemShut {NoStop}%
\bibitem [{\citenamefont {Bemfica}\ \emph {et~al.}(2019{\natexlab{a}})\citenamefont {Bemfica}, \citenamefont {Disconzi},\ and\ \citenamefont {Noronha}}]{Bemfica:2019cop}%
  \BibitemOpen
  \bibfield  {author} {\bibinfo {author} {\bibfnamefont {F.~S.}\ \bibnamefont {Bemfica}}, \bibinfo {author} {\bibfnamefont {M.~M.}\ \bibnamefont {Disconzi}}, \ and\ \bibinfo {author} {\bibfnamefont {J.}~\bibnamefont {Noronha}},\ }\href {\doibase 10.1103/PhysRevLett.122.221602} {\bibfield  {journal} {\bibinfo  {journal} {Phys. Rev. Lett.}\ }\textbf {\bibinfo {volume} {122}},\ \bibinfo {pages} {221602} (\bibinfo {year} {2019}{\natexlab{a}})},\ \Eprint {http://arxiv.org/abs/1901.06701} {arXiv:1901.06701 [gr-qc]} \BibitemShut {NoStop}%
\bibitem [{\citenamefont {Kovtun}(2019)}]{Kovtun:2019hdm}%
  \BibitemOpen
  \bibfield  {author} {\bibinfo {author} {\bibfnamefont {P.}~\bibnamefont {Kovtun}},\ }\href {\doibase 10.1007/JHEP10(2019)034} {\bibfield  {journal} {\bibinfo  {journal} {JHEP}\ }\textbf {\bibinfo {volume} {10}},\ \bibinfo {pages} {034} (\bibinfo {year} {2019})},\ \Eprint {http://arxiv.org/abs/1907.08191} {arXiv:1907.08191 [hep-th]} \BibitemShut {NoStop}%
\bibitem [{\citenamefont {Bemfica}\ \emph {et~al.}(2019{\natexlab{b}})\citenamefont {Bemfica}, \citenamefont {Bemfica}, \citenamefont {Disconzi}, \citenamefont {Disconzi}, \citenamefont {Noronha},\ and\ \citenamefont {Noronha}}]{Bemfica:2019knx}%
  \BibitemOpen
  \bibfield  {author} {\bibinfo {author} {\bibfnamefont {F.~S.}\ \bibnamefont {Bemfica}}, \bibinfo {author} {\bibfnamefont {F.~S.}\ \bibnamefont {Bemfica}}, \bibinfo {author} {\bibfnamefont {M.~M.}\ \bibnamefont {Disconzi}}, \bibinfo {author} {\bibfnamefont {M.~M.}\ \bibnamefont {Disconzi}}, \bibinfo {author} {\bibfnamefont {J.}~\bibnamefont {Noronha}}, \ and\ \bibinfo {author} {\bibfnamefont {J.}~\bibnamefont {Noronha}},\ }\href {\doibase 10.1103/PhysRevD.100.104020} {\bibfield  {journal} {\bibinfo  {journal} {Phys. Rev. D}\ }\textbf {\bibinfo {volume} {100}},\ \bibinfo {pages} {104020} (\bibinfo {year} {2019}{\natexlab{b}})},\ \bibinfo {note} {[Erratum: Phys.Rev.D 105, 069902 (2022)]},\ \Eprint {http://arxiv.org/abs/1907.12695} {arXiv:1907.12695 [gr-qc]} \BibitemShut {NoStop}%
\bibitem [{\citenamefont {Hoult}\ and\ \citenamefont {Kovtun}(2020)}]{Hoult:2020eho}%
  \BibitemOpen
  \bibfield  {author} {\bibinfo {author} {\bibfnamefont {R.~E.}\ \bibnamefont {Hoult}}\ and\ \bibinfo {author} {\bibfnamefont {P.}~\bibnamefont {Kovtun}},\ }\href {\doibase 10.1007/JHEP06(2020)067} {\bibfield  {journal} {\bibinfo  {journal} {JHEP}\ }\textbf {\bibinfo {volume} {06}},\ \bibinfo {pages} {067} (\bibinfo {year} {2020})},\ \Eprint {http://arxiv.org/abs/2004.04102} {arXiv:2004.04102 [hep-th]} \BibitemShut {NoStop}%
\bibitem [{\citenamefont {Bemfica}\ \emph {et~al.}(2021)\citenamefont {Bemfica}, \citenamefont {Disconzi}, \citenamefont {Hoang}, \citenamefont {Noronha},\ and\ \citenamefont {Radosz}}]{Bemfica:2020xym}%
  \BibitemOpen
  \bibfield  {author} {\bibinfo {author} {\bibfnamefont {F.~S.}\ \bibnamefont {Bemfica}}, \bibinfo {author} {\bibfnamefont {M.~M.}\ \bibnamefont {Disconzi}}, \bibinfo {author} {\bibfnamefont {V.}~\bibnamefont {Hoang}}, \bibinfo {author} {\bibfnamefont {J.}~\bibnamefont {Noronha}}, \ and\ \bibinfo {author} {\bibfnamefont {M.}~\bibnamefont {Radosz}},\ }\href {\doibase 10.1103/PhysRevLett.126.222301} {\bibfield  {journal} {\bibinfo  {journal} {Phys. Rev. Lett.}\ }\textbf {\bibinfo {volume} {126}},\ \bibinfo {pages} {222301} (\bibinfo {year} {2021})},\ \Eprint {http://arxiv.org/abs/2005.11632} {arXiv:2005.11632 [hep-th]} \BibitemShut {NoStop}%
\bibitem [{\citenamefont {Bemfica}\ \emph {et~al.}(2022)\citenamefont {Bemfica}, \citenamefont {Disconzi},\ and\ \citenamefont {Noronha}}]{Bemfica:2020zjp}%
  \BibitemOpen
  \bibfield  {author} {\bibinfo {author} {\bibfnamefont {F.~S.}\ \bibnamefont {Bemfica}}, \bibinfo {author} {\bibfnamefont {M.~M.}\ \bibnamefont {Disconzi}}, \ and\ \bibinfo {author} {\bibfnamefont {J.}~\bibnamefont {Noronha}},\ }\href {\doibase 10.1103/PhysRevX.12.021044} {\bibfield  {journal} {\bibinfo  {journal} {Phys. Rev. X}\ }\textbf {\bibinfo {volume} {12}},\ \bibinfo {pages} {021044} (\bibinfo {year} {2022})},\ \Eprint {http://arxiv.org/abs/2009.11388} {arXiv:2009.11388 [gr-qc]} \BibitemShut {NoStop}%
\bibitem [{\citenamefont {Hoult}\ and\ \citenamefont {Kovtun}(2022)}]{Hoult:2021gnb}%
  \BibitemOpen
  \bibfield  {author} {\bibinfo {author} {\bibfnamefont {R.~E.}\ \bibnamefont {Hoult}}\ and\ \bibinfo {author} {\bibfnamefont {P.}~\bibnamefont {Kovtun}},\ }\href {\doibase 10.1103/PhysRevD.106.066023} {\bibfield  {journal} {\bibinfo  {journal} {Phys. Rev. D}\ }\textbf {\bibinfo {volume} {106}},\ \bibinfo {pages} {066023} (\bibinfo {year} {2022})},\ \Eprint {http://arxiv.org/abs/2112.14042} {arXiv:2112.14042 [hep-th]} \BibitemShut {NoStop}%
\bibitem [{\citenamefont {Weickgenannt}(2023)}]{Weickgenannt:2023btk}%
  \BibitemOpen
  \bibfield  {author} {\bibinfo {author} {\bibfnamefont {N.}~\bibnamefont {Weickgenannt}},\ }\href {\doibase 10.1103/PhysRevD.108.076011} {\bibfield  {journal} {\bibinfo  {journal} {Phys. Rev. D}\ }\textbf {\bibinfo {volume} {108}},\ \bibinfo {pages} {076011} (\bibinfo {year} {2023})},\ \Eprint {http://arxiv.org/abs/2307.13561} {arXiv:2307.13561 [nucl-th]} \BibitemShut {NoStop}%
\bibitem [{\citenamefont {Rocha}\ \emph {et~al.}(2021)\citenamefont {Rocha}, \citenamefont {Denicol},\ and\ \citenamefont {Noronha}}]{Rocha:2021zcw}%
  \BibitemOpen
  \bibfield  {author} {\bibinfo {author} {\bibfnamefont {G.~S.}\ \bibnamefont {Rocha}}, \bibinfo {author} {\bibfnamefont {G.~S.}\ \bibnamefont {Denicol}}, \ and\ \bibinfo {author} {\bibfnamefont {J.}~\bibnamefont {Noronha}},\ }\href {\doibase 10.1103/PhysRevLett.127.042301} {\bibfield  {journal} {\bibinfo  {journal} {Phys. Rev. Lett.}\ }\textbf {\bibinfo {volume} {127}},\ \bibinfo {pages} {042301} (\bibinfo {year} {2021})},\ \Eprint {http://arxiv.org/abs/2103.07489} {arXiv:2103.07489 [nucl-th]} \BibitemShut {NoStop}%
\bibitem [{\citenamefont {Rocha}\ and\ \citenamefont {Denicol}(2021)}]{Rocha:2021lze}%
  \BibitemOpen
  \bibfield  {author} {\bibinfo {author} {\bibfnamefont {G.~S.}\ \bibnamefont {Rocha}}\ and\ \bibinfo {author} {\bibfnamefont {G.~S.}\ \bibnamefont {Denicol}},\ }\href {\doibase 10.1103/PhysRevD.104.096016} {\bibfield  {journal} {\bibinfo  {journal} {Phys. Rev. D}\ }\textbf {\bibinfo {volume} {104}},\ \bibinfo {pages} {096016} (\bibinfo {year} {2021})},\ \Eprint {http://arxiv.org/abs/2108.02187} {arXiv:2108.02187 [nucl-th]} \BibitemShut {NoStop}%
\bibitem [{\citenamefont {De~Groot}(1980)}]{DeGroot:1980dk}%
  \BibitemOpen
  \bibfield  {author} {\bibinfo {author} {\bibfnamefont {S.~R.}\ \bibnamefont {De~Groot}},\ }\href@noop {} {\emph {\bibinfo {title} {{Relativistic Kinetic Theory. Principles and Applications}}}},\ edited by\ \bibinfo {editor} {\bibfnamefont {W.~A.}\ \bibnamefont {Van~Leeuwen}}\ and\ \bibinfo {editor} {\bibfnamefont {C.~G.}\ \bibnamefont {Van~Weert}}\ (\bibinfo {year} {1980})\BibitemShut {NoStop}%
\bibitem [{\citenamefont {Weickgenannt}\ \emph {et~al.}(2021{\natexlab{b}})\citenamefont {Weickgenannt}, \citenamefont {Speranza}, \citenamefont {Sheng}, \citenamefont {Wang},\ and\ \citenamefont {Rischke}}]{Weickgenannt:2020aaf}%
  \BibitemOpen
  \bibfield  {author} {\bibinfo {author} {\bibfnamefont {N.}~\bibnamefont {Weickgenannt}}, \bibinfo {author} {\bibfnamefont {E.}~\bibnamefont {Speranza}}, \bibinfo {author} {\bibfnamefont {X.-l.}\ \bibnamefont {Sheng}}, \bibinfo {author} {\bibfnamefont {Q.}~\bibnamefont {Wang}}, \ and\ \bibinfo {author} {\bibfnamefont {D.~H.}\ \bibnamefont {Rischke}},\ }\href {\doibase 10.1103/PhysRevLett.127.052301} {\bibfield  {journal} {\bibinfo  {journal} {Phys. Rev. Lett.}\ }\textbf {\bibinfo {volume} {127}},\ \bibinfo {pages} {052301} (\bibinfo {year} {2021}{\natexlab{b}})},\ \Eprint {http://arxiv.org/abs/2005.01506} {arXiv:2005.01506 [hep-ph]} \BibitemShut {NoStop}%
\bibitem [{\citenamefont {Hu}(2022{\natexlab{b}})}]{Hu:2022lpi}%
  \BibitemOpen
  \bibfield  {author} {\bibinfo {author} {\bibfnamefont {J.}~\bibnamefont {Hu}},\ }\href {\doibase 10.1103/PhysRevD.106.036004} {\bibfield  {journal} {\bibinfo  {journal} {Phys. Rev. D}\ }\textbf {\bibinfo {volume} {106}},\ \bibinfo {pages} {036004} (\bibinfo {year} {2022}{\natexlab{b}})},\ \Eprint {http://arxiv.org/abs/2202.07373} {arXiv:2202.07373 [hep-ph]} \BibitemShut {NoStop}%
\bibitem [{\citenamefont {Denicol}\ \emph {et~al.}(2012)\citenamefont {Denicol}, \citenamefont {Niemi}, \citenamefont {Molnar},\ and\ \citenamefont {Rischke}}]{Denicol:2012cn}%
  \BibitemOpen
  \bibfield  {author} {\bibinfo {author} {\bibfnamefont {G.~S.}\ \bibnamefont {Denicol}}, \bibinfo {author} {\bibfnamefont {H.}~\bibnamefont {Niemi}}, \bibinfo {author} {\bibfnamefont {E.}~\bibnamefont {Molnar}}, \ and\ \bibinfo {author} {\bibfnamefont {D.~H.}\ \bibnamefont {Rischke}},\ }\href {\doibase 10.1103/PhysRevD.85.114047} {\bibfield  {journal} {\bibinfo  {journal} {Phys. Rev. D}\ }\textbf {\bibinfo {volume} {85}},\ \bibinfo {pages} {114047} (\bibinfo {year} {2012})},\ \bibinfo {note} {[Erratum: Phys.Rev.D 91, 039902 (2015)]},\ \Eprint {http://arxiv.org/abs/1202.4551} {arXiv:1202.4551 [nucl-th]} \BibitemShut {NoStop}%
\bibitem [{\citenamefont {Wagner}(2024)}]{Wagner:2024rbt}%
  \BibitemOpen
  \bibfield  {author} {\bibinfo {author} {\bibfnamefont {D.}~\bibnamefont {Wagner}},\ }\emph {\bibinfo {title} {{Quantum kinetic theory and dissipative spin hydrodynamics}}},\ \href {\doibase 10.21248/gups.83488} {Ph.D. thesis},\ \bibinfo  {school} {Frankfurt U.} (\bibinfo {year} {2024})\BibitemShut {NoStop}%
\bibitem [{\citenamefont {Rocha}\ \emph {et~al.}(2022)\citenamefont {Rocha}, \citenamefont {Ferreira}, \citenamefont {Denicol},\ and\ \citenamefont {Noronha}}]{Rocha:2022fqz}%
  \BibitemOpen
  \bibfield  {author} {\bibinfo {author} {\bibfnamefont {G.~S.}\ \bibnamefont {Rocha}}, \bibinfo {author} {\bibfnamefont {M.~N.}\ \bibnamefont {Ferreira}}, \bibinfo {author} {\bibfnamefont {G.~S.}\ \bibnamefont {Denicol}}, \ and\ \bibinfo {author} {\bibfnamefont {J.}~\bibnamefont {Noronha}},\ }\href {\doibase 10.1103/PhysRevD.106.036022} {\bibfield  {journal} {\bibinfo  {journal} {Phys. Rev. D}\ }\textbf {\bibinfo {volume} {106}},\ \bibinfo {pages} {036022} (\bibinfo {year} {2022})},\ \Eprint {http://arxiv.org/abs/2203.15571} {arXiv:2203.15571 [nucl-th]} \BibitemShut {NoStop}%
\bibitem [{\citenamefont {Hess}(2015)}]{Hess:2015szz}%
  \BibitemOpen
  \bibfield  {author} {\bibinfo {author} {\bibfnamefont {S.}~\bibnamefont {Hess}},\ }\href {\doibase 10.1007/978-3-319-12787-3} {\emph {\bibinfo {title} {{Tensors for Physics}}}},\ Undergraduate Lecture Notes in Physics\ (\bibinfo  {publisher} {Springer},\ \bibinfo {year} {2015})\BibitemShut {NoStop}%
\bibitem [{\citenamefont {Panda}\ \emph {et~al.}(2021)\citenamefont {Panda}, \citenamefont {Dash}, \citenamefont {Biswas},\ and\ \citenamefont {Roy}}]{Panda:2020zhr}%
  \BibitemOpen
  \bibfield  {author} {\bibinfo {author} {\bibfnamefont {A.~K.}\ \bibnamefont {Panda}}, \bibinfo {author} {\bibfnamefont {A.}~\bibnamefont {Dash}}, \bibinfo {author} {\bibfnamefont {R.}~\bibnamefont {Biswas}}, \ and\ \bibinfo {author} {\bibfnamefont {V.}~\bibnamefont {Roy}},\ }\href {\doibase 10.1007/JHEP03(2021)216} {\bibfield  {journal} {\bibinfo  {journal} {JHEP}\ }\textbf {\bibinfo {volume} {03}},\ \bibinfo {pages} {216} (\bibinfo {year} {2021})},\ \Eprint {http://arxiv.org/abs/2011.01606} {arXiv:2011.01606 [nucl-th]} \BibitemShut {NoStop}%
\bibitem [{\citenamefont {Barnett}(1935)}]{Barnett:1935wyv}%
  \BibitemOpen
  \bibfield  {author} {\bibinfo {author} {\bibfnamefont {S.~J.}\ \bibnamefont {Barnett}},\ }\href {\doibase 10.1103/RevModPhys.7.129} {\bibfield  {journal} {\bibinfo  {journal} {Rev. Mod. Phys.}\ }\textbf {\bibinfo {volume} {7}},\ \bibinfo {pages} {129} (\bibinfo {year} {1935})}\BibitemShut {NoStop}%
\end{thebibliography}%
\end{document}